\documentclass[letterpaper, unpublished]{quantumarticle}
\pdfoutput=1 

\usepackage[utf8]{inputenc}
\usepackage[english]{babel}
\usepackage[T1]{fontenc}
\usepackage{amsmath}
\usepackage{hyperref}

\usepackage{tikz}
\usepackage{lipsum}

\usepackage{graphicx}
\usepackage{amssymb} 

\begin{document}
\title{A modified Lindblad equation for a Rabi driven electron-spin qubit with tunneling to a Markovian lead}
\author{Emily Townsend}
\email[]{emily.townsend@nist.gov}
\affiliation{ Joint Quantum Institute and Nanoscale Device Characterization Division (Associate), University of Maryland, College Park, Maryland 20742, USA; and National Institute of Standards and Technology, Gaithersburg, Maryland 20899-8423, USA}
\orcid{0000-0001-5669-1048}
\author{Joshua Pomeroy}
\affiliation{Nanoscale Device Characterization Division, National Institute of Standards and Technology, Gaithersburg, Maryland 20899-8423, USA}
\orcid{0000-0001-7848-7179}
\author{Garnett W. Bryant}
\affiliation{Nanoscale Device Characterization Division and Joint Quantum Institute, National Institute of Standards and Technology, Gaithersburg, Maryland 20899-8423, USA; and University of Maryland, College Park, Maryland 20742, USA}
\orcid{0000-0002-2232-0545}

\date{January 22, 2026}

\begin{abstract}
We derive a modified Lindblad equation for the state of quantum dot tunnel coupled to a Markovian lead when the spin state of the dot is driven by an oscillating magnetic field.  We show that the equation is a completely positive, trace-preserving map and find the jump operators. This is a driven-dissipative regime in which coherent driving is relevant to the  tunneling and cannot be treated as simply a rotation modifying the system with a bath derived  under a static  magnetic field. This work was motivated by an experimental desire to determine the Zeeman splitting of an electron spin on a quantum dot (a spin qubit), and in a related work we show that this splitting energy can be found by measuring the charge occupancy of the dot while sweeping the frequency of the driving field \cite{Townsend2025}.  Here we cover the full derivation of the equation and give the jump operators.  These jump operators are potentially useful for describing the stochastic behavior of more complex systems with coherent driving of a spin capable of tunneling on or off of a device, such as in electron spin resonance scanning tunneling microscopy.  The jump operators have the interesting feature of combining jumps of electrons onto and off of the device.
\end{abstract}
\maketitle

\section{\label{Introduction}Introduction}

The quest for quantum information processing and associated technological developments have brought us to the point at which we can create physical devices with truly quantum behavior.  We can now create physical systems that represent qubits, initialize them, perform quantum logic gates on them, and even make rudimentary quantum computations. 
  This means we routinely create in our labs systems whose time evolution consists of nearly coherent, unitary dynamics from the realm of quantum mechanics, in combination with stochastic processes, both intentional and unintentional.  This is the realm of open quantum systems: quantum systems coupled to an external environment.
The ability to design any such system, characterize it once we build it, and to operate it for information processing all rely on our ability to describe the dynamics of an open quantum system (OQS).  In this work we derive the OQS dynamics of a system of just four levels, a very modest size.  But because this system is both driven by a time-dependent field and coupled to a thermal reservoir, this calculation requires us to reexamine the assumptions that ordinarily go into the derivation of a master equation. 

The spin of an unpaired electron confined to a solid-state device, Zeeman split by a static magnetic field, can play the role of a qubit for quantum information processing. A tunnel junction from the device to a lead provides for spin preparation and measurement by Elzerman readout \cite{Elzerman2004}, in which the chemical potential of the lead is tuned to allow one spin state to tunnel off the device, but not the other. 
  An alternating current (AC) magnetic field can be used to manipulate the spin state of the electron via application of a microwave pulse on one of the confinement gates \cite{Koppens2006}.  
  The former is a stochastic, time-irreversible process, while the latter is a unitary process.  Often the stochastic and unitary processes are constrained to separate phases of operation, however some possible advantages may be gained by applying them simultaneously.  In this work we derive differential equations that describe the dynamics of a quantum dot in which the electron is tunnel coupled to a lead while undergoing Rabi oscillations driven by an AC magnetic field.  In a related work, we apply these equations to show that the dynamics and steady state charge occupancy of the system are markedly different depending on whether the  driving field is resonant with the Zeeman-split spin states \cite{Townsend2025}.  This implies that  monitoring the charge occupancy while sweeping the frequency of the AC driving field could be used to experimentally determine the Zeeman splitting.  This is important because the success of all gate manipulations on a spin qubit depend on knowing this splitting and driving at that frequency.  

In this work we derive an equation for the dynamics of the density matrix of the dot from the equations describing the dynamics of the composite system by tracing over the possible states of the lead the electron can tunnel to, along with assumptions about the correlation functions and dynamics in the lead (Markovian approximation, secular approximation).   This largely follows a standard method outlined, for example, in references \cite{Lidar2020, Rivas2012, Breuer2002}.  Generally, when following the standard method, the result is the Gorini-Kossakowski-Sudarshan-Lindblad (GKSL) master equation \cite{Lindblad1976, Gorini1976}, which results in a time evolution that is completely positive and trace-preserving (CPTP).  However, the standard result applies when the system Hamiltonian is time-independent.  The presence of the AC drive means that the system Hamiltonian does not have stationary states, which prevents us from performing an energy spectral decomposition of the interaction-picture operators.  Instead, different-time interaction-picture creation and destruction operators have greatly complicated commutation relations, which prevent us from easily moving from a Markovian approximation to the GKSL equation via the secular approximation. Therefore we evaluate the commutators of time-evolved creation and destruction operators explicitly in matrix form, which then allows us to perform the secular approximation.  This allows us to derive the dissipative part of the dynamics for the system density matrix.  The result is still a CPTP map that can be written in Lindblad form.  The non-dissipative dynamics induced by the coupling to the environment are not covered  here but may be the subject of a future publication.

This work is related to and differs from some earlier  works in the literature.  In references \cite{Martin2003, Engel2001, Engel2002} the hypothesized physical  setup is similar, but it is assumed that the driving field does not affect the tunneling of electrons on  and off of the dot. They obtain a Lindblad equation in the absence of a driving field to describe the tunneling, and then add  on a driving field to the time evolution of the spin state.  This is applicable when there is some other source of decoherence  of the spin state, stronger than that induced by the tunneling.  However,  since the tunneling acts as a measurement of the spin state via Elzerman readout,
when the spin state has a coherent evolution it is more appropriate  to include that evolution when deriving  the Lindblad equation.
Including the driving and making a secular approximation is  similar to the dressed atom description of  resonance fluorescence \cite{Cohen-Tannoudji1977, Cohen-Tannoudji2008, Elouard2020}, and results in a similar splitting of levels as seen in the Autler-Townes (AC Stark) effect \cite{Autler1955}.  The difference is that the dissipative system in this  case causes changes in  the number of electrons, not just in the spin state of the electrons, so the system has more accessible states, i.e. different charge states  as well as different spin states.	
Reference \cite{Mozgunov2020} states that they have derived a completely positive master equation for arbitrary driving via a coarse-graining procedure (which plays a similar role to and, in some limit, reduces to the secular approximation). 
The results here differ substantially from what we would get if we applied the method of reference \cite{Mozgunov2020} to this system.  In that work, each of the jump operators in the Lindblad equation is coarse-grained individually.
In this work, we make a single secular approximation of the entire equation of motion, which also results in a completely positive master equation, however under a substantially different approximation or averaging, which we believe to have greater relevance to the system we are investigating.

\subsection{Structure of the paper}
In the remainder of this introductory section we will describe the system and environment we have in mind and give the Hamiltonian that describes them (subsection \ref{subs:Ham}).  In section \ref{sec:DeriveRho} we will derive a system master equation that follows from this description.  This will involve the standard steps of moving to the interaction picture, expanding the expression for the time-derivative of the full density matrix to second order and making the Born and Markov approximations.  Then in subsection \ref{subs:Secular} the secular approximation is made, in which the coherent drive due to the AC field introduces a significant departure from the standard method.  Subsection \ref{subs:Sums} finishes the derivation by performing the sums and integrals to yield differential equations for the system density matrix.  (The full set of equations is relegated to Appendix G.)  Section \ref{sec:JumpOps} then examines these equations for a particular environment condition: a chemical potential and temperature relevant to the experiment of interest.  We determine jump operators for this condition representing a completely positive and trace preserving map.  Finally, the conclusion (section \ref{sec:Conclusion}) briefly summarizes, makes a comparison to the dressed state picture and speculates on possible extensions of, the work.  Throughout the paper, we have moved more pedagogical parts of the argument and very long or detailed equations to the appendices.

\subsection{System, Environment and Interaction Hamiltonian}\label{subs:Ham}

We wish to describe the dynamics of a system, such as a quantum dot, which can hold a varying number of electrons, and a lead held at an electrostatic potential in thermodynamic equilibrium, which will serve as the environment.  Electrons can tunnel between the lead and the dot at a rate that is slow compared to the magnetic field-induced dynamics of the spin and to the re-equilibration of the environment after a tunnel event. 
We discuss a particular device which motivated this work in greater detail in reference \cite{Townsend2025}.
We will restrict our attention to a submanifold of the system that is relevant for tunneling and assume the dot may have $N-1$, $N$, or $N+1$ electrons on it, where these three charge states are those whose chemical potential, $\mu_D(N) = E_N-E_{N-1}$, is closest to the chemical potential of the tunnel-coupled lead,  $\mu_L $. Assuming $N$ is odd, we map these charge states onto: an empty dot($N-1\rightarrow 0$), a dot with a single electron($N\rightarrow 1$) and one with two electrons, in a singlet state ($N+1 \rightarrow 2$). We identify the energies of these states as $E_{N-1} \equiv 0$, $E_N = E_1$,  and $E_{N+1}=E_b   = 2E_N + U$, where the difference between subsequent electrons increases due to a stong Coulomb interaction, $U= e^2/C >> E_N$, between electrons confined to the same dot with capacitance $C$  \footnote{These energies are measured relative to the $N-1$ energy state, and assume a constant externally applied electric potential, $\phi_{ext}$  contributing an energy $-q\phi_{ext}$ for $q$ the total charge on the dot and an interaction energy $q^2/2C$. }, \cite{Beenakker1991}. 

A static magnetic field, $B_0 \hat{z}$ will lead to a Zeeman splitting of the two spin states of the singly-occupied state by an energy  $\hbar \omega_0 = g\mu_B B_0$, where $g$ is the electron g-factor, $\mu_B$ is the Bohr magneton, and $\hbar$ is the reduced Planck constant. An applied AC magnetic field, $B_1(\hat x\cos \omega t  )$, in the  rotating wave approximation (RWA) becomes $B_1(\hat x\cos \omega t  - \hat y \sin \omega t )$ and  drives rotations in the spin state. The resulting system Hamiltonian (with  $\hbar \omega_1 = g\mu_B B_1$) is 
\begin{align}
H_S(t)= (E_1- \frac{\hbar \omega_0}{2}) \hat{n}_\downarrow +(E_1 + \frac{\hbar \omega_0}{2}) \hat{n}_\uparrow
+ U \hat{n}_\uparrow\hat{n}_\downarrow \nonumber \\
+ \frac{\hbar\omega_1}{2} \left(e^{-i\omega t} \hat a^\dagger_\uparrow \hat a_\downarrow  +  e^{i\omega t} \hat a^\dagger_\downarrow \hat a_\uparrow\right).
\label{eq:SysHam}
 \end{align}

The operators $\hat a^\dagger_s, \hat a_s$ create and destroy an electron of spin $s$ on the dot, while  $\hat{n}_s =\hat a^\dagger_s \hat a_s $ counts the number of electrons in spin state $s $ of the dot. This Hamiltonian is time-dependent, meaning it lacks energy eigenstates.  It does have a unitary time evolution operator,  $\mathcal U(t,t_0)$, which describes evolution from initial time $t_0$ to time $t>t_0$. This is defined by the  equation 
\begin{align}
 i\hbar \frac{\partial}{\partial t} \mathcal U(t,t_0) =  H_S(t)\, \mathcal U(t,t_0)
\end{align}
 which can be found by a series of frame transformations detailed in appendix \ref{App:Frame}, and for which  $(\mathcal U(t,t_0))^\dagger = \mathcal U(t_0,t)$. 

This device interacts with an environment by electrons tunneling to and from the lead.  The total Hamiltonian of the system and environment, $H(t) = H_S(t) +H_E + H_I$ includes contributions from the system Hamiltonian above, the time-independent environment Hamiltonian, $H_E$, for the lead, and the electron-spin preserving, time-independent, interaction Hamiltonian:
 \begin{equation}
H_{I}=  \sum_l \lambda_{l,\uparrow} \hat a^\dagger_{l,\uparrow} \hat a_\uparrow + \lambda_{l,\downarrow} \hat a ^\dagger_{l,\downarrow} \hat a_\downarrow +\lambda_{l,\uparrow} \hat a^\dagger_\uparrow \hat a_{l,\uparrow} + \lambda_{l,\uparrow} \hat a ^\dagger_\downarrow \hat a_{l,\downarrow}
\end{equation}
in which $a^\dagger_{l,s}, a_{l,s}$ create and destroy an electron of spin $s$ in orbital state $l$ in the lead, and $\lambda_{l,s}$ is the tunneling strength between the state with spin $s$ in the dot and the state $l,s$ in the lead, arising from the overlap of orbitals. We will eventually take the approximation that the overlap  is independent of $l$ and $s$, i.e. $\lambda_{l,s} \rightarrow \lambda $.  
 This overlap is small, making $H_I$ weaker than the other two terms of the Hamiltonian.

\section{Derivation of driven-dissipative dynamics}\label{sec:DeriveRho}
In this section we derive the equations describing the dynamics of our spin system.
The first part of this section follows the standard method for deriving a GKSL equation.  The difference from this method arising from the time-dependent Hamiltonian shows up when we carry out a secular approximation, in section \ref{subsubssubs:Structure}.

We wish to understand how the evolution of the density matrix for the system is changed by the presence of the tunnel coupling with the lead.
Beginning with the von Neumann equation for the density matrix of the system and the environment, 
\begin{align}
\dot{\rho}_{S+E}(t) = -\frac{i}{ \hbar}[H(t),\rho_{S+E}(t)]
\label{eq:vN}
\end{align}
we move into the interaction picture with respect to the system and environment Hamiltonians, $H_S$ and $H_E$, to isolate just the change in the evolution of the density matrix that results from the interaction between system and environment.  Operators in the interaction picture have a tilde and evolve according to:
\begin{align}
\tilde{\mathcal{O}}(t) = \mathcal{U}(t_0,t) e^{i H_E t/ \hbar} \mathcal{O} e^{-i H_E t/ \hbar}\mathcal{U}(t,t_0),
\end{align}
where the time dependence of the system Hamiltonian is accounted for by the unitary operator $\mathcal{U}(t,t_0)$, while the time-independent lead Hamiltonian yields a time evolution $e^{-i H_E t/ \hbar}$.
In the interaction picture, the von Neumann equation for the density matrix of the combined dot plus the lead, $\tilde{\rho}_{S+E}$, becomes (see Appendix \ref{App:IntPic})
\begin{align}
\dot{\tilde{\rho}}_{S+E}(t) =-\frac{ i}{\hbar}[\tilde{H}_I(t),\tilde\rho_{S+E}(t)],
\label{eq:vN1}
\end{align}
which has the formal solution (at time $t_1$, later than the initial condition at time $t_0$)
\begin{align}
\tilde{\rho}_{S+E}(t_1) = \tilde{\rho}_{S+E}(t_0) -\frac{i}{ \hbar}\int_{t_0}^{t_1} dt_2 [\tilde{H}_I(t_2),\tilde{\rho}_{S+E}(t_2)].
\label{eq:vN2}
\end{align}
Substituting Eq. \ref{eq:vN2} into Eq. \ref{eq:vN1} yields
\begin{align}
\dot{\tilde{\rho}}_{S+E}(t_1) =& -\frac{i}{ \hbar} [\tilde{H}_I(t_1),\tilde{\rho}_{S+E}(t_0)] \nonumber \\
&+\left( -\frac{i}{ \hbar}\right)^2 \int_{t_0}^{t_1} dt_2 [\tilde{H}_I(t_1) ,[\tilde{H}_I(t_2),\tilde{\rho}_{S+E}(t_2)]].
\label{eq:vN2ndO}
\end{align}
We want to describe the dynamics of the system, rather than the system and the environment.  Soon we will discuss the approximations we can make when the system and environment are only weakly coupled that allow us to reasonably do this.  For now let us consider the dynamics of a density matrix that describes the dot, averaged over every many-body eigenstate of the lead, $|\phi_i^M\rangle_E$, weighted with the probability that the lead is in that state, $\langle \phi_i^M | \tilde{\rho}_E |\phi_i^M\rangle_E$.  The lead is held at some chemical potential, $\mu_L$, and temperature, T, so these states include all possible numbers of electrons, $M$, in the lead. This average is referred to as tracing the combined system and environment density matrix over the lead. 
The time derivative of this system-only density matrix in the interaction picture after taking this trace is:
\begin{align}
\dot{\tilde{\rho}}(t_1) = {\rm Tr}_E\{\dot{\tilde{\rho}}_{S+E}(t_1) \} \equiv \sum_{i,M}\langle \phi_i^M|_E \, \dot{\tilde{\rho}}_{S+E}(t_1) |\phi_i^M\rangle_E.
\label{eq:TraceEnv}
\end{align}
We can now substitute  Eq. \ref{eq:vN2ndO}  into Eq. \ref{eq:TraceEnv}.  Appendix \ref{App:1stTerm0} shows that when we take the trace of Eq. \ref{eq:vN2ndO}, the first term (with only one commutator) is zero because the thermal state of the lead is diagonal in the basis of many-electron states, and states with different numbers of electrons are orthogonal.  

\subsection{Born and Markov approximations}
Up to this point, the dynamics of the system density matrix is unitary.  Hamiltonian dynamics can yield only unitary evolution, and we have assumed that a Hamiltonian describes both the system, the environment and their interaction.  However, a quantum system interacting with a large thermal system is well described as behaving irreversibly.  We will make several assumptions about the system, the bath, and the weakness of their interaction that will result in the appearance of irreversible dynamics.  

The Born approximation consists of the argument that the environment is much larger than, and is only weakly coupled to the system, such that the environment state is unchanged as a function of time, 
 leaving no entanglement between the system and the lead, such that the two are separable: $\tilde{\rho}_{S+E}(t_2) = \tilde{\rho}(t_2) \otimes \rho_{E}$.  While the system's entanglement with its environment causes irreversible behavior in the system, the environment is affecting the system far more than the system is affecting the environment such that we can reasonably approximate the environment as unchanging. 

Because we have expressed the system density matrix in the interaction picture, and the bath coupling is a weak perturbation of the system's time evolution, $\tilde{\rho}$ is not changing very quickly (we've shown in Appendix \ref{App:1stTerm0}  the  term of $\dot{\tilde{\rho}}$  that is first order in $\lambda_{l,s}$ is zero).  The Markovian approximation is that  $\tilde\rho$ is nearly constant over the course of the integral  from $t_0$ to $t_1$, and we may replace $\tilde\rho(t_2)$ with $\tilde\rho(t_1)$, so that the change in the density matrix at $t_1$ is a function of the state of the system at only $t_1$, not the full history of the state of the system at times between $t_0$ and $t_1$.  This yields:

\begin{align}
\dot{\tilde{\rho}}(t_1) 
= \frac{-1}{ \hbar^2} \int_{t_0}^{t_1} dt_2 {\rm Tr}_E  \left( \left[\tilde{H}_I(t_1) ,[\tilde{H}_I(t_2),\tilde{\rho}(t_1) \otimes \rho_{E}] \right] \right).
\label{eq:vNBM}
\end{align}

While Eq. \ref{eq:vN2ndO} could be expanded to every order in $\lambda_{l,s}$ by continued substitution, the Markovian approximation in Eq. \ref{eq:vNBM} terminates this series at second order, and introduces the memoryless property: Now the change in the density matrix at $t_1$ depends only the state of system at $t_1$, and not on all earlier states as it did before.  

The product state of the system and environment density matrices in Eq. \ref{eq:vNBM} allows us to rewrite the trace over the environment in terms of the environment's correlation functions, which we define as:
\begin{align}
\mathrm{N}_{ls}(t_1, t_2) \equiv {\rm Tr}_E \left( \tilde{a}_{l,s}^\dagger (t_1)  \tilde{a}_{l,s} (t_2)    \rho_{E}  \right)\\
\mathrm{G}_{ls}(t_1, t_2) \equiv {\rm Tr}_E \left( \tilde{a}_{l,s} (t_1)  \tilde{a}_{l,s}^\dagger (t_2)    \rho_{E}  \right)
\end{align}
The correlation function $\mathrm{N}_{ls}(t_1, t_2)$  represents the probability that the environment's thermal state is such that an electron could tunnel out of the level with quanum number $l$ and spin $s$ at time $t_2$ and then tunnel back to that level at time $t_1$ (i.e. that level is occupied), transformed into the interaction picture.  The  correlation function $\mathrm{G}_{ls}(t_1, t_2)$ instead measures the probability that the level is unoccupied, allowing an electron to tunnel in first, then back out.
These correlation functions have only one spin index, although the interaction Hamiltonian also generates terms with e.g. $a_{l,\uparrow}^\dagger (t_1) a_{l,\downarrow} (t_2) \rho_E$.  These mixed-spin correlations are zero for a thermal environment of non-interacting electrons in a Gibbs state, though we must keep track of similar coherences for the dot states. For an environment of non-interacting electrons the correlation functions depend only on the difference between the time arguments, $t_1 - t_2$, and certain expectation values (e.g. ${\rm Tr}_E \left\{ \tilde{a}_{l,s} (t_1)  \tilde{a}_{l,s} (t_2)  \rho_{E}  \right\}$ and ${\rm Tr}_E \{ \tilde{a}^\dagger_{l,s} (t_1)  \tilde{a}_{l',s'} (t_2)  \rho_{E} \}$ for $l\neq l'$ or $s \neq s'$)  are zero.  See Appendix \ref{App:ThermalLead}.

Along with a change of variable, $\tau = t_1 - t_2$, $\mathrm{d}\tau = -\mathrm{d}t_2$ (the minus sign disappears once we reverse the direction of integration from the smaller limit to the larger limit), this yields:
\begin{widetext}
\begin{align}
\dot{\tilde{\rho}}(t_1) 
= \frac{-1}{ \hbar^2} \sum_{l,s=\uparrow, \downarrow}  \int_{0}^{t_1- t_0} d\tau \,\,
& \lambda_{l,s}^2  \left[ \tilde{a}_s(t_1),   \tilde{a}^\dagger_s(t_1-\tau) \tilde{\rho}(t_1 )  \right]
\mathrm{N}_{l,s}(\tau) 
 + \lambda_{l,s}^2 \left[ \tilde{a}^\dagger_s(t_1),  \tilde{a}_s(t_1-\tau)  \tilde{\rho}(t_1)  \right]
\mathrm{G}_{l,s}(\tau)
\nonumber \\
- & \lambda_{l,s}^2  \left[ \tilde{a}_s(t_1) ,  \tilde{\rho}(t_1) \tilde{a}^\dagger_s(t_1) \right]
\mathrm{G}_{l,s}(-\tau) 
- \lambda_{l,s}^2 \left[
\tilde{a}^\dagger_s(t_1)  , \tilde{\rho}(t_1)\tilde{a}_s(t_1-\tau) \right]
\mathrm{N}_{l,s}(-\tau). 
\label{eq:vNGNTau}
\end{align}
\end{widetext}
where we have used the invariance of cyclic permutation of the trace.

\subsection{Secular approximation}\label{subs:Secular}

The secular approximation is that we will not be able to observe very rapid fluctuations in the system density matrix as they average to zero.  

In order to obtain an irreversible description from a unitary evolution, some information contained in the unitary evolution must be lost somewhere.  That may happen in a bath which regains its thermal equilibrium (as in the Born approximation) or because we don't have the ability to track very small (Markov approximation) or high frequency (secular  approximation) changes in our system, so that up to the ability of our measurements, information is lost to us.  

For greater accuracy, the secular approximation could be replaced by a coarse-graining of the evolution of the system with a timestep 
 $T_A$ which is much larger than the timescale at which the environment can regain equilibrium after a perturbation (the correlation time of the environment, $\tau_c$), and much smaller than the timescale on which the system density matrix is changing due to interactions with the environment (the right hand side of Eq. \ref{eq:vNBM}).  
 In the limit of $T_A \rightarrow \infty$, coarse-graining is equivalent to the secular approximation \cite{Lidar2001, Lidar2020, Mozgunov2020}, which is sufficient for our purposes. 
 
\subsubsection{Structure due to non-commuting time evolved creation and destruction operators}\label{subsubssubs:Structure}
It is in carrying out the secular approximation 
that we need to address the fact that the unitary time evolution due to the system Hamiltonian has more structure than typical.  This prevents us from simply writing the creation and destruction operators in the frequency domain by expanding in a basis of energy eigenstates, as is usually done (e.g. Eq. 519 of reference \cite{Lidar2020}).  While the different-spin fermionic anticommutator $\{ a_{l,\uparrow}, a^\dagger_{l, \downarrow}\}$ is zero, the unitary time evolution mixes spin up and spin down so $\{ \tilde{a}_{l,\uparrow}(t_1), \tilde{a}^\dagger_{l, \downarrow}(t_2)\} = \{ \mathcal{U}(0,t_1) \tilde{a}_{l,\uparrow} \, \mathcal{U}(t_1,0), \mathcal{U}(0,t_2)\tilde{a}^\dagger_{l, \downarrow}\, \mathcal{U}(t_2,0)\}$ is not zero, and is, in fact, a function of both the times $t_1$ and $t_2$, and not only their difference.

In order to accommodate this structure, we use Matlab\footnote{MATLAB 2025b, The MathWorks, Inc., Natick, Massachusetts, United States.  Mention of commercial products is for information only; it does not imply recommendation or endorsement by NIST.} 
to perform the necessary matrix multiplications in equation \ref{eq:vNGNTau}. The script which performs the multiplications and makes appropriate substitutions is available as supplemental information (LindbladSubstitutionScriptForPub.m and its raw and formatted output, LindbladSubstitutions.tex and  LindbladSubstitutions.pdf) \cite{supp}. We obtain the time evolution of each matrix element of the reduced system as the sum of multiple terms, yielding a set of coupled differential equations for all of the components of the system density matrix, $\tilde{\rho}_{\alpha,\beta}(t)  = \langle \alpha | \tilde{\rho}(t) |  \beta \rangle= \langle \alpha |\mathcal{U}(0,t)  \rho \,\mathcal{U}(t,0) | \beta \rangle$, where $\alpha$ and $\beta$ are any of the states of the system: unoccupied ($0$); singly occupied, spin up ($\uparrow$);  singly occupied, spin down ($\downarrow$); or doubly occupied in a singlet state (b).    Appendix \ref{App:MatrixMult} shows the relevant of these equations before carrying out the secular approximation.  
Because the system Hamiltonian preserves the number of electrons on the dot but not the spin state of the dot, the coherences of the density matrix between states with different numbers of electrons (e.g. $\tilde{\rho}_{0,\uparrow}$) do not couple with the diagonal elements of the density matrix (populations), and will simply decay away if they are not initially zero, while coherences between different spin states ($\tilde{\rho}_{\uparrow, \downarrow}$) do couple to the diagonal elements of the density matrix (Eq. \ref{eq:UpDnCouples}).  The full set of equations for all matrix elements is given in the supplemental information \cite{supp}.

The various functions of time that appear in the elements of $\mathcal{U}$ (see Eq. \ref{eq:U}) appear as products in the integral in Eq. \ref{eq:vNGNTau}, and we perform a secular approximation on each term (product).  For example,
with $\mathcal{B}(t,t_0) \equiv \langle \uparrow | \mathcal{U} (t, t_0) | \downarrow \rangle$, also defined in equation \ref{eq:U}, and  $\Omega \equiv \sqrt{\omega_1^2 + (\omega - \omega_0)^2}$ the generalized Rabi frequency,
  some of the terms in Eq. \ref{eq:vNGNTau} contain the product (where the overline indicates a complex conjugate):
\begin{align}
 \mathcal{B}&(0,t_1-\tau) \overline{\mathcal{B}(0,t_1)}   \nonumber \\
&=\left( \frac{\omega_1}{2\Omega}\right)^2 e^{-iE_1 \tau/\hbar}e^{i (\omega/2) \tau}  \left(e^{i(\Omega/2) \tau} (1  -e^{-i\Omega t_1}) 
\right. \nonumber \\ &  \left.
\qquad \qquad \qquad \qquad \qquad+ e^{-i(\Omega/2) \tau}(1- e^{i \Omega t_1}) \right)  \nonumber \\
&\approx\left( \frac{\omega_1}{2\Omega}\right)^2 e^{-iE_1 \tau /\hbar}e^{i (\omega/2) \tau}  \left(e^{i(\Omega/2) \tau} + e^{-i(\Omega/2) \tau}  \right).
\label{eq:BB*}
\end{align}
	The last line is the secular approximation: that the rapid oscillations (at frequency  $\pm \Omega$) in time $t_1$ average to zero over timescales we observe.  The variable $t_1$ is the argument of the density matrix and its time derivative in equation \ref{eq:vNGNTau}. Averaging or coarse-graining over this variable is applicable when $t_1 >> \hbar/( \Delta E)> \tau_c$, where $\Delta E$ is the transition energy represented by this term and  $\tau_c \sim \hbar /(k_{\rm B} T)$ with $k_{\rm B}$ the Boltzman constant and $T$ the temperature, is the correlation time of the environment (discussed in Appendix \ref{App:ThermalLead})\cite{Lidar2020}.  In this example $\Delta E = 0- (\frac{E_1}{\hbar} - \frac{\omega}{2} - \frac{\Omega}{2})$, a transition between the empty dot and the lowest dressed level occupied by an electron.
The variable $\tau$ is the integration variable and may be arbitrarily small, so we do not average over the terms containing it.

After performing the secular approximation for each such product, we obtain differential equations for the system density matrix, $\tilde{\rho}(t_1)$, at time $t_1$ with no dependence on $t_1$ other than that present in the system density matrix, i.e. a time-independent map.  
We show here the equation for just a single component of the density matrix, after taking $\lambda_{l,s} = \lambda$:
\begin{widetext}
\begin{align}
 -\frac{d\tilde{\rho}_{\uparrow \uparrow}(t_1)}{dt} = \sum_l \frac{\lambda^2}{\hbar^2} \int_0^{t_1 -t_0} d\tau \, & \mathrm{\tilde{\rho}_{\uparrow \uparrow}}(t_1)
\left( 
\mathrm{G}_{l,\uparrow}(-\tau) \left( \left(  \frac{\Omega-a}{2\Omega} \right)^2 e^{-i\omega_{+-}\tau}  + \left( \frac{\Omega+a}{2\Omega} \right)^2   e^{-i\omega_{++}(\tau)} \right)
\right. \nonumber \\
&\qquad \left.
+\overline{\mathrm{N}_{l,\uparrow}(-\tau)}e^{-iE_b\tau/\hbar}\left( \frac{\omega_1}{2\Omega}\right)^2  \left(e^{i\omega_{--} \tau} + e^{i\omega_{-+} \tau}  \right)
\right. \nonumber \\
&\qquad \left.
+  \overline{\mathrm{G}_{l,\downarrow}(-\tau)} \left( \frac{\omega_1}{2\Omega}\right)^2  \left(e^{i\omega_{--} \tau} + e^{i\omega_{-+} \tau}  \right)
\right. \nonumber \\
&\qquad \left.
+\mathrm{N}_{l,\downarrow}(-\tau)e^{iE_b\tau/\hbar}\left( \left(  \frac{\Omega-a}{2\Omega} \right)^2 e^{-i\omega_{+-}\tau}  + \left( \frac{\Omega+a}{2\Omega} \right)^2   e^{-i\omega_{++}(\tau)}\right)
+\mathrm{c.c.}
\right) \nonumber \\
+& \mathrm{\tilde{\rho}_{\uparrow \downarrow}}(t_1)
\left(
\left(\mathrm{G}_{l,\uparrow}(-\tau) +e^{iE_b\tau/\hbar}\mathrm{N}_{l,\downarrow}(-\tau)\right)
\left( -\frac{\omega_1}{2\Omega} \right) \left(\frac{\Omega-a}{2\Omega}e^{-i\omega_{+-}\tau}  - \frac{\Omega+a}{2\Omega} e^{-i\omega_{++}\tau}\right)
\right. \nonumber \\
&\qquad \left.
+\left( e^{iE_b\tau/\hbar}\mathrm{N}_{l,\uparrow}(-\tau) +\mathrm{G}_{l,\downarrow}(-\tau) \right) \left(  -\frac{\omega_1}{2\Omega} \right)
\left( \frac{\Omega+a}{2\Omega}e^{-i\omega_{--}\tau}- \frac{\Omega-a}{2\Omega}e^{-i\omega_{-+}\tau} \right) 
\right)
 \nonumber \\
+& \mathrm{\tilde{\rho}_{\downarrow \uparrow}}(t_1)
\left(  
\left( \overline{\mathrm{G}_{l,\uparrow}(-\tau)} + e^{-iE_b\tau/\hbar} \overline{\mathrm{N}_{l,\downarrow}(-\tau)} \right) 
\left( -\frac{\omega_1}{2\Omega} \right) \left(\frac{\Omega-a}{2\Omega}e^{i\omega_{+-}\tau}  -\frac{\Omega+a}{2\Omega}e^{i\omega_{++}\tau}\right)
 \right. \nonumber \\
&\qquad \left. 
+\left(e^{-iE_b\tau/\hbar} \overline{\mathrm{N}_{l,\uparrow}(-\tau)} +  \overline{\mathrm{G}_{l,\downarrow}(-\tau)} \right)
\left(  -\frac{\omega_1}{2\Omega} \right)
\left( \frac{\Omega+a}{2\Omega}e^{i\omega_{--}\tau}- \frac{\Omega-a}{2\Omega}e^{i\omega_{-+}\tau} \right)
\right)
 \nonumber \\
-& \mathrm{\tilde{\rho}_{00}}(t_1)
\left(
+\overline{\mathrm{N}_{l,\uparrow}(-\tau)}\left( \left(  \frac{\Omega-a}{2\Omega} \right)^2 e^{-i\omega_{+-}\tau}  + \left( \frac{\Omega+a}{2\Omega} \right)^2   e^{-i\omega_{++}(\tau)}  \right)
 \right. \nonumber \\
&\qquad \left. 
+\mathrm{N}_{l,\downarrow}(-\tau) \left( \frac{\omega_1}{2\Omega}\right)^2  \left(e^{i\omega_{--} \tau} + e^{i\omega_{-+} \tau}  \right)
+\mathrm{c.c.}
\right)
 \nonumber \\
-& \mathrm{\tilde{\rho}_{bb}}(t_1)
\left(
\mathrm{G}_{l,\uparrow}(-\tau)e^{-iE_b\tau/\hbar} \left( \frac{\omega_1}{2\Omega}\right)^2  \left(e^{i\omega_{--} \tau} + e^{i\omega_{-+} \tau}  \right) 
 \right. \nonumber \\
&\qquad \left. 
+  \overline{\mathrm{G}_{l,\downarrow}(-\tau)}e^{iE_b\tau/\hbar} \left( \left(  \frac{\Omega-a}{2\Omega} \right)^2 e^{-i\omega_{+-}\tau}  + \left( \frac{\Omega+a}{2\Omega} \right)^2   e^{-i\omega_{++}(\tau)}  \right)
+\mathrm{c.c.}
\right)
\end{align}
\end{widetext}
where $a= \omega - \omega_0$, $\Omega = \sqrt{\omega_1^2 + a^2}$, $E_b   = 2E_1 + U$, and $$ \omega_{--} \equiv \frac{E_1}{\hbar} - \frac{\omega}{2} - \frac{\Omega}{2};$$
$$ \omega_{-+} \equiv \frac{E_1}{\hbar} - \frac{\omega}{2}+ \frac{\Omega}{2};$$
$$\omega_{+-} \equiv \frac{E_1}{\hbar} + \frac{\omega}{2} - \frac{ \Omega}{2};$$
$$ \omega_{++} \equiv \frac{E_1}{\hbar} + \frac{\omega}{2}+\ \frac{\Omega}{2}.$$

\subsection{Performing the sum and integration}\label{subs:Sums}
We can now carry out the sum over states in the bath ($l,s$) and the integral over $\tau$.  Given the environment's rapid decay back to thermal equilibrium ($\tau_c << t_1 - t_0$), the environment's correlation functions, $\mathrm{G}_{l,s}(\tau)$ and $\mathrm{N}_{l,s}(\tau)$, are zero for $\tau >> \tau_c$, allowing us to replace the upper limit of integration over $\tau$ with infinity, and make use of the half-time Fourier transform of the bath correlation functions and the associated principal parts integrals \cite{Dutra2004}:

\begin{align}
\Gamma_{ls} (\omega) &\equiv \int_{0}^{\infty} \mathrm{d}t  \, \mathrm{G}_{ls}(t) e^{-i\omega t} =\int_{0}^{\infty} \mathrm{d}t  \, \overline{\mathrm{G}_{ls}(-t)} e^{-i\omega t}\\
G_{ls} (\omega) &\equiv \int_{-\infty}^{\infty} \mathrm{d}t  \, \mathrm{G}_{ls}(t) e^{-i\omega t}    = 2 {\rm Re} \{ \Gamma_{ls}(\omega) \} \\
i\sigma_{ls} (\omega) &\equiv   \frac{i}{\pi}\int_{-\infty}^{\infty} \mathrm{d}\omega' \mathcal{P}\frac{1}{\omega' - \omega} G_{ls}(\omega') 
\nonumber \\
& =\int_{-\infty}^{\infty} \mathrm{d}t \,\mathrm{sgn}(t)  \, \mathrm{G}_{ls}(t)  
 = 2i \mathrm{Im} \{ \Gamma_{ls}(\omega) \}
\end{align}

\begin{align}
\nu_{ls} (\omega) &\equiv \int_{0}^{\infty} \mathrm{d}t  \, \mathrm{N}_{ls}(t) e^{-i\omega t} \\
N_{ls} (\omega) &\equiv \int_{-\infty}^{\infty} \mathrm{d}t  \, \mathrm{N}_{ls}(t) e^{-i\omega t}   
  = 2 {\rm Re} \{ \nu_{ls}(\omega) \} \\
i\zeta_{ls} (\omega) &\equiv   \frac{i}{\pi}\int_{-\infty}^{\infty} \mathrm{d}\omega' \mathcal{P}\frac{1}{\omega' - \omega} N_{ls}(\omega') 
 \nonumber \\
& =\int_{-\infty}^{\infty} \mathrm{d}t \,\mathrm{sgn}(t)  \, \mathrm{N}_{ls}(t) 
 = 2i \mathrm{Im} \{ \nu_{ls}(\omega) \} 
\end{align}
\begin{widetext}
Replacing the time-integrated bath correlation functions with the relevant real and imaginary parts of the Fourier transformed functions  yields coupled differential equations for the elements of the system density matrix. 
We show only one here, the rest appear in appendix \ref{App:ResultsGN}:

\begin{align}
-\frac{d\tilde{\rho}_{\uparrow \uparrow}}{dt}= \frac{\lambda^2}{\hbar^2} \sum_l  & \mathrm{\tilde{\rho}_{\uparrow \uparrow}}
\left( 
\left(  \frac{\Omega-a}{2\Omega} \right)^2 G_{l\uparrow}(-\omega_{+-})  + \left( \frac{\Omega+a}{2\Omega} \right)^2   G_{l\uparrow}(-\omega_{++})
+\left( \frac{\omega_1}{2\Omega}\right)^2  \left(N_{l\uparrow}(\frac{E_b}{\hbar}-\omega_{--}) + N_{l\uparrow}(\frac{E_b}{\hbar}-\omega_{-+} )  \right)
\right. \nonumber \\
&\qquad \left.  
+ \left( \frac{\omega_1}{2\Omega}\right)^2  \left( G_{l\downarrow}(-\omega_{--}) + G_{l\downarrow}(-\omega_{-+})  \right)
+ \left(  \frac{\Omega-a}{2\Omega} \right)^2 N_{l\downarrow}(\frac{E_b}{\hbar}-\omega_{+-})  + \left( \frac{\Omega+a}{2\Omega} \right)^2   N_{l\downarrow}(\frac{E_b}{\hbar}-\omega_{++}) 
\right) \nonumber \\
+& \mathrm{\tilde{\rho}_{\uparrow \downarrow}}
\left( -\frac{\omega_1}{2\Omega} \right)
 \left(\frac{\Omega-a}{2\Omega}
\left( \overline{\Gamma_{l\uparrow}(-\omega_{+-})} +\overline{\nu_{l\downarrow}(\frac{E_b}{\hbar}-\omega_{+-})} 
- \overline{\nu_{l\uparrow}(\frac{E_b}{\hbar}-\omega_{-+})} -\overline{\Gamma_{l\downarrow}(-\omega_{-+})} \right)
\right. \nonumber \\
& \qquad \qquad \qquad \left. 
-\frac{\Omega+a}{2\Omega}
\left( \overline{\Gamma_{l\uparrow}(-\omega_{++})} +\overline{\nu_{l\downarrow}(\frac{E_b}{\hbar}-\omega_{++})}
-
  \overline{\nu_{l\uparrow}(\frac{E_b}{\hbar}-\omega_{--})}  - \overline{\Gamma_{l\downarrow}(-\omega_{--})} \right) 
 \right) 
 \nonumber \\
+& \mathrm{\tilde{\rho}_{\downarrow \uparrow}}
\left(
\left( -\frac{\omega_1}{2\Omega} \right) \left( \frac{\Omega-a}{2\Omega} 
\left( \Gamma_{l\uparrow}(-\omega_{+-}) +\nu_{l\downarrow}(\frac{E_b}{\hbar}-\omega_{+-}) 
-\nu_{l\uparrow}(\frac{E_b}{\hbar}-\omega_{-+}) -\Gamma_{l\downarrow}(-\omega_{-+})
 \right)
 \right. \right. \nonumber \\
&\qquad \qquad \qquad \left. \left.
  + 
\frac{\Omega+a}{2\Omega} \left( \nu_{l\uparrow}(\frac{E_b}{\hbar}-\omega_{--}) +\Gamma_{l\downarrow}(-\omega_{--}) 
- \Gamma_{l\uparrow}(-\omega_{++}) -\nu_{l\downarrow}(\frac{E_b}{\hbar}-\omega_{++})
\right)    \right)
\right)
 \nonumber \\
-& \mathrm{\tilde{\rho}_{00}}
\left(
 \left(  \frac{\Omega-a}{2\Omega} \right)^2 N_{l\uparrow}(\omega_{+-})  + \left( \frac{\Omega+a}{2\Omega} \right)^2   N_{l\uparrow}(\omega_{++})  
+\left( \frac{\omega_1}{2\Omega}\right)^2   \left( N_{l\downarrow}(\omega_{--}) + N_{l\downarrow}(\omega_{-+})  \right)
\right)
 \nonumber \\
-& \mathrm{\tilde{\rho}_{bb}}
\left(
 \left( \frac{\omega_1}{2\Omega}\right)^2  \left( G_{l\uparrow}(\omega_{--}-\frac{E_b}{\hbar}) + G_{l\uparrow}(\omega_{-+}-\frac{E_b}{\hbar})  \right) 
\right.
\nonumber \\
&\qquad \left. 
+   \left(  \frac{\Omega-a}{2\Omega} \right)^2 G_{l\downarrow}(\omega_{+-}-\frac{E_b}{\hbar})  + \left( \frac{\Omega+a}{2\Omega} \right)^2   G_{l\downarrow}(\omega_{++}-\frac{E_b}{\hbar})  
\right)
\label{eq:rhoupupGNGammaNu}
\end{align}

 \end{widetext}
The Fourier transformed bath correlation functions, $G(\omega)$ and $N(\omega)$ are the real part of complex functions ($\Gamma(\omega), \zeta(\omega)$) which have imaginary parts ($\sigma(\omega), \nu(\omega)$).  The real and imaginary parts of the right hand side of Eq. \ref{eq:rhoupupGNGammaNu}  contribute  to the dissipative and oscillatory parts of the interaction picture dynamics, respectively.  The dissipative part describes tunneling (changes in the number of electrons on the dot) while the oscillatory part changes the spin state of an electron on the dot.  The oscillatory part is not limited to Lamb shifts in the Schrodinger picture, because of the nature of the time evolution operator used to enter the interaction picture.  This oscillatory behavior is in addition to and smaller than the Rabi oscillations that will reappear when transforming back from the interaction picture  (because the environment interaction is weak compared to the AC field driving). Therefore we limit discussion to the dissipative parts of these equations in this work, and leave a discussion of the oscillatory part for a future work.

Assuming an ordinary (non-superconducting) lead in a thermal state with temperature $T$ and chemical potential $\mu$ and with a density of states $D(\hbar \omega)$ (which we will assume is constant over the width of the band and zero elsewhere), Appendix \ref{App:ThermalLead} shows 
\begin{align}
\sum_l N_{ls} (\hbar \omega) 
= D(\hbar \omega) n_e(\hbar \omega, \mu, T) 2\pi \hbar
\end{align}
\begin{align}
\sum_l G_{ls} (\hbar \omega)
= D(-\hbar \omega) n_h(-\hbar \omega, \mu, T) 2\pi \hbar
\end{align}
in which $ n_e(\hbar \omega, \mu, T)$ represents the probability that an electron occupies the state at an energy $\hbar \omega$ above the chemical potential $\mu$ of the lead at temperature $T$, and $n_h$ similarly for holes.

Upon performing these sums over $l$, and assuming the density of states is constant, $D(\hbar \omega) = D,$  we obtain the final results, which are given in full detail in appendix G, and one of which is shown here (with the variables $\mu$ and $T$ omitted from the Fermi functions for brevity):

\begin{widetext}
\begin{align}
-\frac{d\tilde{\rho}_{\uparrow \uparrow}}{dt}= \frac{2\pi \lambda^2 D}{\hbar^2}  & \mathrm{\tilde{\rho}_{\uparrow \uparrow}}
\left( 
\left(  \frac{\Omega-a}{2\Omega} \right)^2 n_h(\hbar \omega_{+-})  + \left( \frac{\Omega+a}{2\Omega} \right)^2   n_h(\hbar \omega_{++})
+\left( \frac{\omega_1}{2\Omega}\right)^2  \left(n_e(E_b- \hbar \omega_{--}) + n_e(E_b- \hbar \omega_{-+} )  \right)
\right. \nonumber \\
&\qquad \left.  
+ \left( \frac{\omega_1}{2\Omega}\right)^2  \left( n_h(\hbar \omega_{--}) + n_h(\hbar \omega_{-+})  \right)
+ \left(  \frac{\Omega-a}{2\Omega} \right)^2 n_e(E_b- \hbar \omega_{+-})  + \left( \frac{\Omega+a}{2\Omega} \right)^2   n_e(E_b- \hbar \omega_{++}) 
\right) \nonumber \\
+& \mathrm{\tilde{\rho}_{\uparrow \downarrow}}
\left( -\frac{\omega_1}{2\Omega} \right)
 \left(\frac{\Omega-a}{2\Omega}
\left( \frac{1}{2}n_h(\hbar \omega_{+-}) +\frac{1}{2}n_e(E_b- \hbar \omega_{+-}) 
- \frac{1}{2}n_e(E_b- \hbar \omega_{-+}) -\frac{1}{2}n_h(\hbar \omega_{-+}) \right)
\right. \nonumber \\
&\qquad  \qquad\left.
-\frac{\Omega+a}{2\Omega}
\left( \frac{1}{2}n_h(\hbar \omega_{++}) +\frac{1}{2}n_e(E_b- \hbar \omega_{++})
-
  \frac{1}{2}n_e(E_b- \hbar \omega_{--}) -\frac{1}{2}n_h(\hbar \omega_{--}) \right) 
 \right) 
 \nonumber \\
+& \mathrm{\tilde{\rho}_{\downarrow \uparrow}}
\left(
\left( -\frac{\omega_1}{2\Omega} \right) \left(\frac{\Omega-a}{2\Omega}
\left( \frac{1}{2}n_h(\hbar \omega_{+-}) +\frac{1}{2}n_e(E_b- \hbar \omega_{+-}) 
-\frac{1}{2}n_e(E_b- \hbar \omega_{-+}) -\frac{1}{2}n_h(\hbar \omega_{-+})
 \right)
\right. \right. \nonumber \\
&\qquad  \qquad \left. \left.
  + 
\frac{\Omega+a}{2\Omega} \left( \frac{1}{2}n_e(E_b- \hbar \omega_{--}) +\frac{1}{2}n_h(\hbar \omega_{--}) 
- \frac{1}{2}n_h(\hbar \omega_{++}) -\frac{1}{2}n_e(E_b- \hbar \omega_{++})
\right)    \right)
\right)
 \nonumber \\
-& \mathrm{\tilde{\rho}_{00}}
\left(
 \left(  \frac{\Omega-a}{2\Omega} \right)^2 n_e(\hbar \omega_{+-})  + \left( \frac{\Omega+a}{2\Omega} \right)^2   n_e(\hbar \omega_{++})  
+\left( \frac{\omega_1}{2\Omega}\right)^2   \left( n_e(\hbar \omega_{--}) +n_e(\hbar \omega_{-+})  \right)
\right)
 \nonumber \\
-& \mathrm{\tilde{\rho}_{bb}}
\left(
 \left( \frac{\omega_1}{2\Omega}\right)^2  \left( n_h(-\hbar \omega_{--}+ E_b) + n_h(-\hbar \omega_{-+} + E_b)  \right) 
\right. \nonumber \\
&\qquad  \qquad\left.
+  \left(  \frac{\Omega-a}{2\Omega} \right)^2 n_h(-\hbar \omega_{+-}+ E_b)  + \left( \frac{\Omega+a}{2\Omega} \right)^2   n_h(-\hbar \omega_{++} + E_b)  
\right)
\label{eq:rhoupupdot}
\end{align}

 \end{widetext}
Notice that the energies that appear in the arguments of the $n_h$ and $n_e$ are transition energies.  The probability of a transition from one state of the system to another is determined by the probability that the environment can make a transition with a similar (opposite) energy difference and particle number (i.e., there is an electron or hole with an energy below or above the chemical potential of the environment). The  ground (no electrons) state of the system has been defined with zero energy, so many of the transitions in Eq. \ref{eq:rhoupupdot} appear as a function of a single energy, though they are properly transitions (i.e. $n_h(\hbar \omega_{+-} - 0)$).

 The assumption of a thermal state in the lead is key to driving the nature of the dissipation in the system.  We have assumed that the lead does not maintain any superposition of states with different numbers of electrons for a time longer than $\tau_c$.  This assumption of rapid decoherence in the bath, along with the Born and secular approximations, is the source of irreversibility in these equations. 
A set of coupled differential equations $\dot{\tilde{\rho}} = M \tilde{\rho}$, exhibit irreversibility if the matrix $M$ has at least one negative eigenvalue, which leads to exponential decay rather than oscillatory behavior. If all eigenvalues are negative there is a unique steady state. In either case, given a late-time state one cannot determine which initial condition one started with, meaning the dynamics are irreversible.  We have not shown that the matrix connecting the full  component equations in appendix \ref{App:Results} has any negative eigenvalues, though it likely does given the physics they describe.  However, we will show in the next section that for a particular set of assumptions on the leads these equations are equivalent to a Lindblad equation, which is always irreversible.

\section{Jump operators and demonstratng that the map is CPTP}\label{sec:JumpOps}

\begin{figure}
\includegraphics[width=0.45\textwidth]{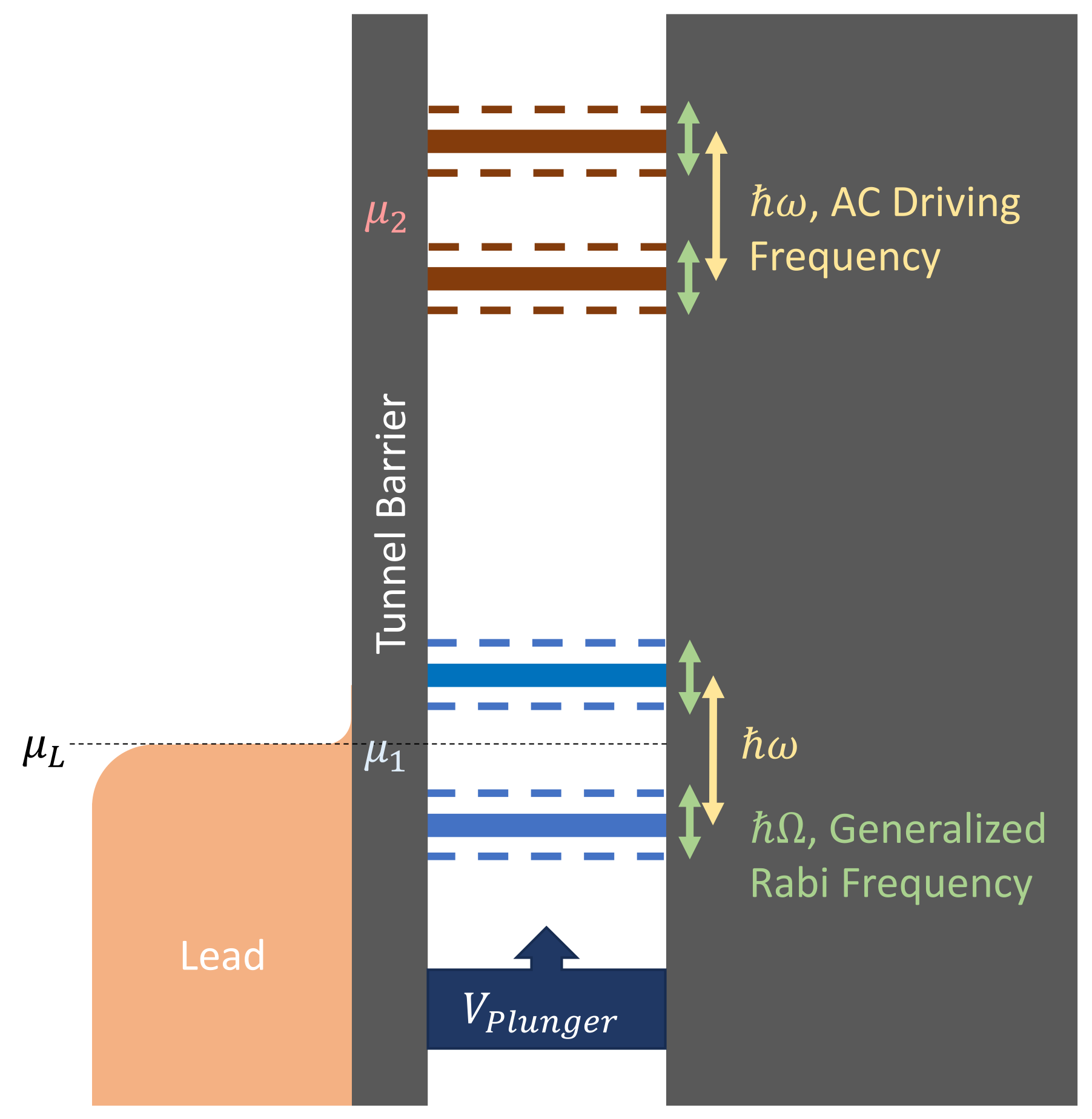}
 \caption{ (from reference  \cite{Townsend2025})  Relative one-electron-addition transition energies of different states of the dot and of the lead (i.e. chemical potentials of the dot and lead). } 
\label{fig:muLevels}
\end{figure}

The full dissipative part of the dynamics of the system density matrix in the interaction picture is given by the coupled differential equations in appendix \ref{App:Results}.  However, their character can be seen more clearly with a few physical assumptions.  We choose the conditions which are explored in reference \cite{Townsend2025} and depicted in Figure \ref{fig:muLevels}, which shows the transition energies for various ways the dot can change its occupancy by a single electron, and how they compare to the chemical potential of the lead.   The chemical potential of the dot (lead) is the change in energy due to changing the number of electrons on the dot (lead) by one, \textit{at equilibrium}.  We assume the lead is at equilibrium for examining the dynamics of the dot, which is not at equilibrium. However, we use the symbol $\mu$ here for all transition energies which change the electron number by one. 
With no magnetic fields, the transition from zero to one electrons increases the energy of the dot by $\mu_1 = E_1 - E_0 = E_1$ (we define  $E_0 = 0$) and from one to two electrons, $\mu_2 = E_b - E_1 = 2E_1 + U - E_1$.
With simply a static magnetic field, $B_0$, only the 1-electron state has an unpaired spin, so spin up and spin down states are split, $E_{\uparrow} = E_1 +  \hbar \omega_0/2$ and $E_{\downarrow} = E_1 - \hbar\omega_0/2$.  Without driving $\mu_1$ is split by the Zeeman energy: $\mu_\uparrow = E_\uparrow$, $\mu_\downarrow = E_\downarrow$, as is $\mu_2$: $\mu_{2\uparrow} = E_b - E_\uparrow$, $\mu_{2\downarrow} = E_b - E_\downarrow$. 
 When there are both static and driving fields, the Markovian and secular approximations to the tunneling create the effect of ``chemical potentials" that are instead split by both the driving frequency and the Rabi frequency.  These are the transition energies that appear in the Fermi occupation functions of Eq. \ref{eq:rhoupupdot}:  $ \hbar\omega_{++} = E_1 +\hbar\omega/2 +  \hbar\Omega/2$, $ \hbar\omega_{+-} =  E_1 +\hbar\omega/2 -  \hbar\Omega/2 $, $ \hbar\omega_{-+} =  E_1 -\hbar\omega/2+  \hbar\Omega/2  $, $ \hbar\omega_{--} =  E_1 -\hbar\omega/2 -  \hbar\Omega/2$, along with $E_b - \hbar \omega_{++}$, $E_b - \hbar \omega_{+-}$, $E_b - \hbar \omega_{-+}$, $E_b - \hbar \omega_{--}$ .  We have so far derived equations for an arbitrary temperature and chemical potential for the lead, and here examine the regime of very low temperature and a chemical potential of the lead between $ E_1 -\hbar\omega/2 +  \hbar\Omega/2$  and $  E_1 +\hbar\omega/2 -  \hbar\Omega/2$, relevant to experiment.  These assumptions simply reflect the fact that these experiments are performed at millikelvin temperatures, and the voltage of the lead will be tuned to be well in the regime such that tunneling off the dot is possible for the spin-up electrons but not spin-down electrons.  These equations apply at any driving magnitude ($\propto \omega_1$), driving frequency ($\omega$), and static field magnitude ($\propto \omega_0$), so long as the tunneling between the lead and the dot ($\hbar/2\pi \lambda^2 D$) is slower than the dynamics on the dot ($2\pi/\Omega$), and both are slower than decoherence in the lead ($\tau_c$).

These conditions have electron and hole occupancies of the lead:
\begin{align}
n_e(\hbar \omega_{--}) = n_e(\hbar\omega_{-+}) =1 \\
 n_h(\hbar\omega_{++}) = n_h(\hbar\omega_{+-}) = 1 \\
n_h(E_b - \hbar\omega_{\rm any}) = 1 \nonumber \\
n_h(\hbar \omega_{--}) = n_h(\hbar \omega_{-+}) =0 \\
 n_e(\hbar \omega_{++}) = n_e(\hbar \omega_{+-}) = 0 \\
n_e(E_b - \hbar \omega_{\rm any}) = 0.
\end{align}
As a reminder, the generalized Rabi frequency is $\Omega = \sqrt{\omega_1^2 + a^2}$ (with detuning $a = \omega- \omega_0$ and $\omega_0 >> \omega_1 \sim \Omega$) and $\omega_{\rm any}$ is any of $\omega_{++}, \, \omega_{+-}, \, \omega_{-+},\, \omega_{--}$.

This leads to the following equations describing the dynamics of the system  interaction picture density matrix:
\begin{widetext}
\begin{align}
-\frac{d\tilde{\rho}_{0,0}}{dt} 
&=\frac{ 2\pi \lambda^2 D}{\hbar} \left(\tilde{\rho}_{00} 
-\frac{1}{2}\left(1+\frac{a^2}{\Omega^2}\right)\tilde{\rho}_{\uparrow \uparrow}
- \frac{\omega_1^2}{2\Omega^2}\tilde{\rho}_{\downarrow \downarrow} 
-  \frac{\omega_1}{2\Omega}\frac{a}{\Omega}\tilde{\rho}_{\downarrow \uparrow}
-  \frac{\omega_1}{2\Omega}\frac{a}{\Omega}\tilde{\rho}_{\uparrow \downarrow} 
\right) \qquad  \text{(off-resonance)} \label{eq:rho00OFF} \\
&=\frac{ 2\pi \lambda^2 D}{\hbar} \left(\tilde{\rho}_{00} 
-\frac{1}{2}\tilde{\rho}_{\uparrow \uparrow}
 - \frac{1}{2}\tilde{\rho}_{\downarrow \downarrow} 
\right)  \qquad \text{(on-resonance)}
\\
&=\frac{ 2\pi \lambda^2 D}{\hbar} \left(\tilde{\rho}_{00} 
-\tilde{\rho}_{\uparrow \uparrow}
\right)  \qquad \mathrm{(no\, AC \,driving)}\\
%
-\frac{d\tilde{\rho}_{\uparrow \uparrow}}{dt}
&= 
\frac{ 2\pi \lambda^2 D}{\hbar} \left(  \frac{1}{2}\left( 1+\frac{a^2}{\Omega^2} \right) \mathrm{\tilde{\rho}_{\uparrow \uparrow}}
 + \frac{\omega_1}{2\Omega} \frac{a}{2\Omega}\mathrm{\tilde{\rho}_{\uparrow \downarrow}}
+\frac{\omega_1}{2\Omega} \frac{a}{2\Omega} \mathrm{\tilde{\rho}_{\downarrow \uparrow}}
-\frac{1}{2}\left(\frac{\omega_1}{\Omega}\right)^2 \mathrm{\tilde{\rho}_{00}}
- \mathrm{\tilde{\rho}_{bb}} 
\right)
\qquad  \text{(off-resonance)}\label{eq:rhoupupOFF} \\
&= 
\frac{ 2\pi \lambda^2 D}{\hbar} \left( \mathrm{\tilde{\rho}_{\uparrow \uparrow}} \frac{1}{2}
- \mathrm{\tilde{\rho}_{00}}
\frac{1}{2}  
- \mathrm{\tilde{\rho}_{bb}}
\right) \qquad  \text{(on-resonance)}
\\
&= 
\frac{ 2\pi \lambda^2 D}{\hbar} \left( \mathrm{\tilde{\rho}_{\uparrow \uparrow}} 
- \mathrm{\tilde{\rho}_{bb}}
\right) \qquad  \mathrm{(no \, AC \, driving)}
\\
%
-\frac{d\tilde{\rho}_{\downarrow, \downarrow }}{dt} 
&=\frac{ 2\pi \lambda^2 D}{\hbar} 
\left(  \frac{2\omega_1^2}{4\Omega^2} \tilde{\rho}_{\downarrow \downarrow}\
+\frac{a \omega_1}{4\Omega^2}\tilde{\rho}_{\uparrow \downarrow} 
	+\frac{a \omega_1}{4\Omega^2}\tilde{\rho}_{\downarrow \uparrow}
	- \frac{1}{2}(1+\frac{a^2}{\Omega^2})\tilde{\rho}_{00}
-\tilde{\rho}_{bb} 
\right) \qquad \text{(off-resonance)}\label{eq:rhodowndownOFF}\\
&=\frac{ 2\pi \lambda^2 D}{\hbar} 
\left(\frac{1}{2}\tilde{\rho}_{\downarrow \downarrow} 
-\frac{1}{2} \tilde{\rho}_{00} 
-\tilde{\rho}_{bb} 
\right)  \qquad \text{(on-resonance)}
\label{eq:ZeroTempDiffEq00}
\\
&=\frac{ 2\pi \lambda^2 D}{\hbar} 
\left(
- \tilde{\rho}_{00} 
-\tilde{\rho}_{bb} 
\right)  \qquad \text{(no  AC driving)}
\\
%
-\frac{d\tilde{\rho}_{b,b}}{dt} 
&=\frac{ 2\pi \lambda^2 D}{\hbar} \,  \left(
2 \tilde{\rho}_{bb}
\right) \qquad \text{(on or off-resonance, with or without driving)}\label{eq:rhobbOFF}
\\
%
-\frac{d \tilde{\rho}_{\uparrow \downarrow}}{dt}
&=
\frac{ 2\pi \lambda^2 D}{\hbar} 
\left(\frac{\tilde{\rho}_{\uparrow \downarrow}}{2}
+\frac{a \omega_1}{4\Omega^2}\tilde{\rho}_{\uparrow \uparrow}
+\frac{a \omega_1}{4\Omega^2}\tilde{\rho}_{\downarrow \downarrow} 
+\frac{a \omega_1}{2\Omega^2}\tilde{\rho}_{00} 
\right)  \qquad \text{(off-resonance)} \label{eq:rhoupdownOFF}
\\
&=
\frac{ 2\pi \lambda^2 D}{\hbar}  
\frac{\tilde{\rho}_{\uparrow \downarrow}}{2}	
 \qquad \text{(on-resonance or without driving)}.
 \label{eq:ZeroTempDiffEqUD}
\end{align}
The off-resonance equations simplify to the resonant driving case by taking $a=0$ and $\Omega=\omega_1$, and to the case without AC driving by taking $\omega_1=0$ and $\Omega=a$.

Using the method of reference \cite{Kasatkin2023} we can show that these equations represent a map that is completely positive and trace preserving and equivalent to the Lindblad equation:
\begin{align} 
\frac{d \tilde{\rho}}{dt} = \mathcal{L}\tilde{\rho}
&= \sum_{\alpha} \gamma_\alpha \left( J_\alpha \tilde{\rho} J_\alpha^\dagger - \frac{1}{2}\left\{ J_\alpha^\dagger J_\alpha, \tilde{\rho} \right\}\right) 
\label{eq:LindbladDiag}
\end{align}
 in which
 $\gamma_{1} =  \gamma_{2} =  2\pi D\lambda^2/\hbar$;
 $ \gamma_{3} = \gamma_{4} =\frac{ 2\pi D\lambda^2}{\hbar}\frac{\Omega+a}{2\,\Omega}$;
  $\gamma_{5} =   \gamma_{6} =\frac{2\pi D\lambda^2}{\hbar}\frac{\Omega-a}{2\,\Omega};$ and

\begin{align}
J_{1} =i |\downarrow \rangle \langle b| 
; \qquad
J_{2}= i |\uparrow \rangle \langle b| 
\end{align}
 
\begin{align}
J_{3} =
\frac{\Omega+a}{2\Omega} |0\rangle \langle\uparrow|
+\frac{\omega _{1}}{2\Omega}  |0\rangle \langle\downarrow|
-\frac{\Omega-a}{2\Omega} |\uparrow\rangle \langle 0|
+\frac{\omega _{1}}{2\Omega}|\downarrow\rangle \langle 0| 
\end{align}
 
\begin{align}J_{4} =
+i\frac{\omega _{1}}{2\Omega}|0\rangle \langle\uparrow|
+i\frac{\Omega-a}{2\Omega} |0\rangle \langle\downarrow|
+i\frac{\omega _{1}}{2\Omega}|\uparrow\rangle \langle 0| 
-i\frac{\Omega+a}{2\Omega}|\downarrow\rangle \langle 0|
\end{align}
 
\begin{align}
J_{5} =
-\frac{\Omega-a}{2\Omega} |0\rangle \langle\uparrow|
+\frac{\omega _{1}}{2\Omega} |0\rangle \langle\downarrow|
+\frac{\Omega+a}{2\Omega} |\uparrow\rangle \langle 0|
+\frac{\omega _{1}}{2\Omega} |\downarrow\rangle \langle 0| 
\end{align}
 
\begin{align}J_{6}=
- i\frac{\omega _{1}}{2\Omega}  |0\rangle \langle\uparrow| 
+ i\frac{\Omega+a}{2\Omega}|0\rangle \langle\downarrow|
- i\frac{\omega _{1}}{2\Omega} |\uparrow\rangle \langle 0|
- i\frac{\Omega-a}{2\Omega} |\downarrow\rangle \langle 0|
\end{align}
 \end{widetext}
 
The jump operator $J_1$  ($J_2$) represents a single electron of spin-up (spin-down) tunneling off the dot, changing the state from the doubly occupied ($N+1$) state to the singly occupied ($N$) state.  The reverse jump onto the  dot does not occur for the temperature ($T=0$) and chemical potential (well below the chemical potential of the $N+1$ state) we have chosen in obtaining these equations.  Unlike the non-driven case at this temperature and chemical potential, it is equally likely that either the spin-up or spin-down electron will tunnel off (leaving behind the electron of the other spin), because the driving field is mixing the destination states ($|\uparrow\rangle$ and $|\downarrow \rangle $).

 However, the other four jump operators are not so obviously intuitive.  They all represent transitions between singly occupied ($N$-electron) states and  unoccupied ($N-1$-electron) states.  But they mix whether it is a spin up or a spin down electron that is tunneling, and whether the electron is tunneling on or off the device.  The coefficients, $\gamma_\alpha$, in Eq. \ref{eq:LindbladDiag} come in doubly degenerate pairs, i.e.  $\gamma_1 = \gamma_2$, $\gamma_3 = \gamma_4$, $\gamma_5 = \gamma_6$, so the diagonal Lindblad equation can be rewritten in terms of linear combinations of $J_1$ and $J_2$, (likewise $J_3$ and $J_4$; and $J_5$ and $J_6$).  When the driving frequency is resonant with the Zeeman splitting ($a=0$, $\Omega=\omega_1$) it is possible to combine jump operators as linear combinations $J_3 \pm i J_4$ and $J_5 \pm i J_6$ such that each jump operator represents only tunneling on or only tunneling off the dot, with the spin up and spin down equally mixed in those jumps.  In this resonant case this is consistent with our intuition of a number transition mixed with spin flips.  However, for an arbitrary driving off-resonance, there is no linear combination that cleanly divides the jump operators into those in which an electron leaves the dot and those in which it enters the dot. 

 The supplemental information  \cite{supp} contains a Matlab script (ThereAndBackAgainForPub.m, and its formatted output, 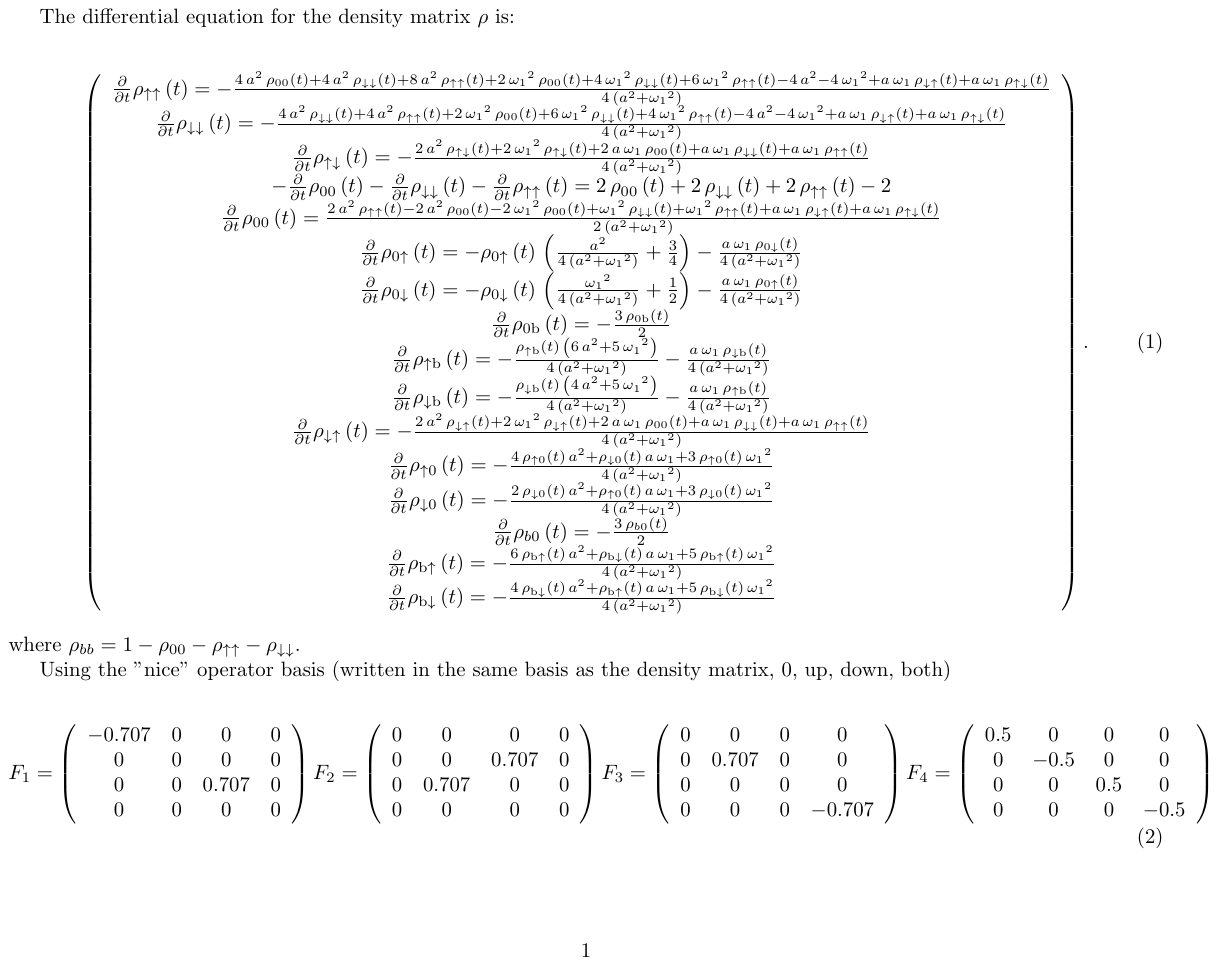) that transforms the differential equations for the elements of the density matrix (Eqs. \ref{eq:rho00OFF}, \ref{eq:rhoupupOFF}, \ref{eq:rhodowndownOFF}, \ref{eq:rhobbOFF}, \ref{eq:rhoupdownOFF}) into an equation for a coherence vector based on a good operator basis (a set of traceless, orthonormal, self-adjoint matrices, $F_m = F_m^\dagger$, $m = 1:15$).   Using the technique from  reference \cite{Kasatkin2023} it then calculates the ``a-matrix'' of the Liouvillian operator written in terms of that operator basis,
 \begin{align}
 \frac{d \tilde{\rho}}{dt} = \mathcal{L}\tilde{\rho}
&= \sum_{m,n} a_{m,n} \left( F_m \tilde{\rho} F_n^\dagger - \frac{1}{2}\left\{F_n^\dagger F_m, \tilde{\rho} \right\}\right). 
\label{eq:Lindbladamatrix}
 \end{align}
  and shows that the eigenvalues of $a$  are all non-negative, meaning that the map is completely positive and trace-preserving (CPTP).  The script then uses the eigenvectors of the a-matrix to transform the jump operators into the diagonal basis, which gives us Eq. \ref{eq:LindbladDiag}. (There are nine additional eigenvalues equal to zero whose corresponding operators do not contribute to the dynamics.)

\section{Conclusion}\label{sec:Conclusion}

We have derived a Lindblad master equation for a Rabi-driven spin in tunneling contact with a thermal lead by explicitly working out the commutation relations for time-evolved particle creation and destruction operators, followed by a time-averaging in the form of the secular approximation. Reference \cite{Townsend2025} shows that this master equation may be evaluated to show that at low temperatures with the chemical potential of the lead midway in energy between the two spin states, the Rabi drive alters the steady-state occupation of the dot.  Specifically, when the Rabi drive is in resonance with the energy splitting between the two spin states, the steady state has an equal probability of being in the spin-up state, the spin-down state, and the state in which the electron has tunneled off the dot.  The charge occupancy of the dot is highly sensitive to whether the drive is resonant, implying that this could be a measurement technique for determining the resonant frequency.

\subsection{Comparison to the dressed state picture of Cohen-Tannoudji, et al.  and Elouard et al.}  These three references \cite{Cohen-Tannoudji1977, Cohen-Tannoudji2008, Elouard2020} and many-more describe a system similar to our spinful electron driven by a microwave field in the dressed state picture, which can then be used to describe the Mollow triplet and the Autler-Townes double resonance.  The dressed state is similar to our transformation into the rotating frame, $U_{RF}$, Eq. \ref{eq:RotFrame}, in which the system Hamiltonian is no longer time dependent.  However, these works have such a system interacting with a bath of photons, not tunneling into a bath of electrons.  In this case the system-bath interaction Hamiltonian can be readily transformed into the rotating frame and has no time-dependence.  However, the tunneling interaction Hamiltonian contains single creation and destruction operators of electrons in the original spin-up and spin-down basis.  This does not transform as readily, which has led us to the method described here.  

\subsection{Future work}
It should be straightforward to extend this work to the case with two reservoirs at different chemical potentials by calculating the relevant jump operators for each reservoir independently using the technique here, and then combining them into a single master equation for the system.  The extension to two reservoirs could inform measurements of current through gated quantum dots \cite{Hanson2007} if an additional driving field were to be added.  In addition, the steady state of the system and the steady state current through the system would be relevant for descriptions of electron spin resonance-scanning tunneling microscopy (ESR-STM)\cite{Baumann2015}.  ESR-STM is currently described using the assumption that the spin system being measured by the STM tip is in thermal contact with three independent baths, one representing tunneling electrons, one representing the microwave field, and one representing interactions with a substrate, as well as the assumption that the average electron occupancy does not change with time (because as many electrons jump on the dot as jump off) \cite{Ast2024}.	  The method described in this work treats the microwave field not as a bath, but as a source of coherent oscillations, and can detail the jumps of individual electrons, allowing a finer description of the spin system interacting with the STM tip.

It is also potentially of interest to extend the work here to the description of more complex spin systems driven by an AC field and tunnel  coupled to a reservoir.  This is less straightforward, because increasing the size of the Hilbert space for the system Hamiltonian will increase the number of terms that must be tracked.  However, if tunneling onto the system can only occur locally, e.g. onto one site at one end of a chain, then the jump operators derived here can readily be applied to the stochastic part of the dynamics while the coherent dynamics are obtained from numerical methods.

\section{Acknowledgements}
ET wishes to acknowledge helpful conversations with Mike Kolodrubetz, Mike Zwolak, and the  support of the National Science Foundation under Grant No. NSF PHY-1748958.
\appendix
\begin{widetext}
\section{Frame Transformations to solve the system Hamiltonian}
\label{App:Frame}

The system Hamiltonian for a spin in a quantum dot is modelled as:
\begin{align}
H_S(t)= (E_1- \frac{\hbar \omega_0}{2}) \hat{n}_\downarrow +(E_1 + \frac{\hbar \omega_0}{2}) \hat{n}_\uparrow
+ U \hat{n}_\uparrow\hat{n}_\downarrow 
+ \frac{\hbar\omega_1}{2} \left(e^{-i\omega t} \hat a^\dagger_\uparrow \hat a_\downarrow  +  e^{i\omega t} \hat a^\dagger_\downarrow \hat a_\uparrow\right).
\label{eq:SysHamApA}
 \end{align}
 If we restrict the Hamiltonian to the single-electron basis (i.e. a two-dimensional basis with only the spin-up and spin down states) 
the restricted Hamiltonian is:
\begin{align}
H_s(t) = E_1 \mathbb{I} + \frac{\hbar\omega_0}{2}\hat{\sigma}_z +\frac{\hbar\omega_1}{2}\left(\cos(\omega t)\hat{\sigma}_x + \sin(\omega t)\hat{\sigma}_y \right),
\end{align}
where $\hat{\sigma}_x$, $\hat{\sigma}_y$, $\hat{\sigma}_z$ are the Pauli spin matrices and $\mathbb{I}$ is a two-by-two identity matrix.  We will solve this Hamiltonian using two subsequent unitary transformations.  The first transformation will be to a frame of reference that is rotating along with the AC field, $B_1$.  In the rotating frame there will be only a static field, which will cause precession characterized by a time evolution operator, $T(t)$.  A second transformation with the adjoint of that time evolution operator, $T^\dagger(t)$, will take us to a frame with no dynamics, i.e. the Hamiltonian in the twice transformed frame will be zero.  In this basis all states are solutions, and any initial state may be time evolved by transforming back to the original reference frame.

The transformation to the rotating frame is a rotation about the z-axis:
\begin{align}
U_{RF}(t) = e^{i\frac{\omega}{2} t\hat{\sigma}_z} = e^{i\frac{\omega}{2} t} |\uparrow\rangle\langle \uparrow | + e^{-i\frac{\omega}{2} t}  |\downarrow \rangle\langle\downarrow |.
\label{eq:RotFrame}
\end{align}
(The second part of the above equality is obtained by expanding the exponential as a power series, and recognizing that $\hat{\sigma}_z^{2n} = \mathbb{I}$ and $\hat{\sigma}_z^{2n+1} = \hat{\sigma}_z$ when $n$ is an  integer, then resumming the $|\uparrow\rangle\langle \uparrow | $ and $ |\downarrow \rangle\langle\downarrow |$ terms separately.)
After this transformation the Hamiltonian is:
\begin{align}
\breve{H} &= U_{RF}(t) H_s(t) U_{RF}^\dagger + i\hbar \frac{\partial U_{RF}}{\partial t}U_{RF}^\dagger
 = U_{RF}(t) H_s(t) U_{RF}^\dagger - i\hbar U_{RF}\frac{\partial U_{RF}^\dagger}{\partial t}
 \nonumber \\
 & =  E_1 \mathbb{I} + \frac{\hbar(\omega_0 - \omega)}{2}\hat{\sigma}_z +\frac{\hbar\omega_1}{2}\hat{\sigma}_x, 
\end{align}
such that if $H_s | \psi(t)\rangle = \frac{i}{\hbar}\frac{\partial}{\partial t} |\psi(t)\rangle$ then $ |\breve{\psi}(t)\rangle = U_{RF}|\psi(t)\rangle$ solves $\breve{H}|\breve{\psi}(t)\rangle =\frac{i}{\hbar}\frac{\partial}{\partial t} |\breve{\psi}(t)\rangle$.

This is now a Hamiltonian with a static field, proportional to $-a\hat{z} +\omega_1\hat{x}$, with $a \equiv \omega- \omega_0$ the detuning of the driving field from the Zeeman energy gap. It will be useful to define the generalized Rabi frequency, $\Omega \equiv  \sqrt{\omega_1^2 + a^2} $ 
 and  $\hat{n}$ as a unit vector that points in the direction of that static field.  The new Hamiltonian can be solved with the transformation:
\begin{align}
\breve{T}^\dagger = e^{it\breve{H}/\hbar} = e^{ iE_1 t/\hbar} e^{i \Omega t \hat{\sigma}_n/2} = e^{ iE_1 t/\hbar}\left( \mathbb{I}\cos(\Omega t/2) + i\hat{\sigma}_n \sin(\Omega t/2)  \right),
\end{align}
where 
\begin{align}
\hat{\sigma}_n =\left(
\begin{array}{cc}
\cos \theta & \sin \theta e^{-i\phi}\\
\sin \theta e^{i \phi} & -\cos \theta
\end{array}\right)
 =\left(
\begin{array}{cc}
\frac{a}{\Omega} & \frac{\omega_1}{\Omega}\\
\frac{\omega_1}{\Omega} & - \frac{a}{\Omega}
\end{array}\right).
\end{align}
Again the exponential of the operator may be expanded and the odd and even powers evaluated as $\hat{\sigma}_n$ and $\mathbb{I}$, respectively.

The twice transformed Hamiltonian is 
\begin{align}
\breve{\breve{H}}=\breve{T}^\dagger (t) \breve{H}  \breve{T}(t)+ i \hbar\frac{\partial \breve{T}^\dagger}{\partial t} \breve{T} = 0
\end{align}
This second transformation has removed all dynamics, and any time-independent state satisfies $\frac{i}{\hbar} \frac{\partial}{\partial t} |\breve{\breve{\psi}} \rangle = \breve{\breve{H}}|\breve{\breve{\psi}}\rangle = 0$.  Thus any initial time-independent state may be evolved forward in time from $t=0$ by reversing the transformations:
\begin{align}
|\psi (t)\rangle &= U_{RF}^\dagger (t) \breve{T}(t) |\psi(0) \rangle
\equiv
\mathbb{U}^\dagger (t) |\psi(0) \rangle
\end{align}
with
\begingroup
\renewcommand*{\arraystretch}{1.5}

\begin{align}
\mathbb{U}^\dagger (t)&=e^{ -iE_1 t/\hbar}\left( 
\begin{array}{cc}
e^{-i\frac{\omega}{2} t}\left(\cos(\Omega t/2) - i\frac{a}{\Omega} \sin(\Omega t/2)\right) &
-i\frac{\omega_1}{\Omega}e^{-i\frac{\omega}{2} t} \sin \Omega t/2 \\
-i\frac{\omega_1}{\Omega}e^{i\frac{\omega}{2} t} \sin \Omega t/2 &
e^{i\frac{\omega}{2} t}\left(\cos(\Omega t/2) +i\frac{a}{\Omega} \sin(\Omega t/2) \right)
\end{array} \right) \nonumber \\
&=e^{ -iE_1 t/\hbar}\left( 
\begin{array}{cc}
e^{-i\frac{\omega}{2} t}\left(\frac{\Omega+a}{2\Omega} e^{-i\Omega t/2}+\frac{\Omega-a}{2\Omega}e^{i\Omega t/2} \right) &
-\frac{\omega_1}{2\Omega}e^{-i\frac{\omega}{2} t} \left( e^{i\Omega t/2 } - e^{-i\Omega t/2} \right) \\
-\frac{\omega_1}{2\Omega}e^{i\frac{\omega}{2} t}  \left( e^{i\Omega t/2 } - e^{-i\Omega t/2} \right)  &
e^{i\frac{\omega}{2} t}\left( \frac{\Omega-a}{2\Omega}e^{-i\Omega t/2}+  \frac{\Omega+a}{2\Omega}e^{i\Omega t/2} \right)
\end{array} \right).
\end{align}
Or most generally, the time evolution from time $t_0$ to $t$ is:
\begin{align}
|\psi(t)\rangle =  U_{RF}^\dagger (t) \breve{T}(t)  \breve{T}^\dagger (t_0) U_{RF} (t_0) |\psi(t_0)\rangle 
=  \mathbb{U}^\dagger (t) \mathbb{U} (t_0) |\psi(t_0)\rangle
\end{align}
\endgroup

Extending this evolution to include the $N-1$ ($|0\rangle$) and $N+1$ ($|b\rangle$) states gives us a unitary time evolution operator written in the basis $|0\rangle, |\uparrow\rangle, |\downarrow \rangle , |b\rangle = |\uparrow \downarrow \rangle$:

\begin{equation}
\mathcal U (t,t_0) = \left(\begin{array}{cccc}
e^{-iE_0(t-t_0)/\hbar}=1 & 0 & 0 & 0 \\
0 & \mathcal{A}(t-t_0) & \mathcal{B}(t,t_0) & 0 \\
0 & \mathcal{C}(t,t_0) & \mathcal{D}(t-t_0) & 0 \\
0 & 0 & 0 & e^{-iE_b(t-t_0)/\hbar}
\end{array}\right)
\label{eq:U}
\end{equation} 
where the diagonal elements depend only on the difference in times, but the off diagonal elements depend on their sum and difference:
$$ \mathcal{A}(t-t_0) = \frac{\Omega-a}{2\Omega}e^{-i\omega_{+-}\tau}+\frac{\Omega+a}{2\Omega}e^{-i\omega_{++}\tau}; $$
$$ \mathcal{B}(t,t_0) = -\frac{\omega_1}{2\Omega}e^{-iE_1\tau}e^{-i(\omega/2)\sigma}(e^{i(\Omega/2) \tau}-e^{-i(\Omega/2)\tau}); $$
$$ \mathcal{C}(t,t_0) = -\frac{\omega_1}{2\Omega}e^{-iE_1\tau}e^{i(\omega/2)\sigma}(e^{i(\Omega/2) \tau}-e^{-i(\Omega/2)\tau}); $$
$$ \mathcal{D}(t-t_0) = \frac{\Omega+a}{2\Omega}e^{-i\omega_{--}\tau}+\frac{\Omega-a}{2\Omega}e^{-i\omega_{-+}\tau}; $$
and
$$\tau \equiv t-t_0;\,\, \sigma \equiv t+t_0$$
$$ \omega_{--} \equiv \frac{E_1}{\hbar} - \frac{\omega}{2} - \frac{\Omega}{2};$$
$$ \omega_{-+} \equiv \frac{E_1}{\hbar} - \frac{\omega}{2}+ \frac{\Omega}{2};$$
$$\omega_{+-} \equiv \frac{E_1}{\hbar} + \frac{\omega}{2} - \frac{ \Omega}{2};$$
$$ \omega_{++} \equiv \frac{E_1}{\hbar} + \frac{\omega}{2}+\ \frac{\Omega}{2}$$
This time evolution operator obeys:
\begin{align}
\frac{-i }{\hbar}H_S\, \mathcal U(t,t_0)&= \frac{\partial}{\partial t} \mathcal U(t,t_0), \label{eq:HU} \\
\frac{i}{\hbar}\mathcal U(t_0,t) H_S &= \frac{\partial}{\partial t} \mathcal U(t_0,t).
\label{eq:UH}
\end{align}
Because $B^*(t,t_0) = \mathcal{C}(t_0,t)$, $(\mathcal U(t,t_0))^\dagger = \mathcal U(t_0,t)$.  Also it is unitary: $\mathcal U^\dagger \mathcal U = \mathcal U \mathcal U^\dagger = \mathbb I$.

\section{Interaction picture}
\label{App:IntPic}

The full Hamiltonian $H = H_S + H_E + H_I$ is the sum of the system, environment and interaction Hamiltonians.  The full system plus environment density matrix satisfies the von Neumann equation:
\begin{align}
\dot{\rho}_{S+E} = \frac{-i}{\hbar}[H_S + H_E + H_I,\rho_{S+E}]
\label{eq:vNAppB}.
\end{align}
The system density matrix is the full density matrix traced over the environment: $\rho_S = {\rm Tr_E}\, \rho_{S+E}$.
We wish to move into the interaction picture 
with respect to the system and environment Hamiltonians, $H_S$ and $H_E$.  
In the interaction picture, operators (with a tilde indicating they are in the interaction picture) evolve according to:
\begin{align}
\tilde{\mathcal{O}}(t) = \mathcal{U}(0,t) e^{i H_E t/\hbar} \mathcal{O} e^{-i H_E t/\hbar}\mathcal{U}(t,0)
\end{align}
which means that the full density matrix in the interaction picture is:
\begin{align}
\tilde{\rho}_{S+E} =   \mathcal{U}(0,t) e^{i H_E t/\hbar} \rho_{S+E}\, e^{-i H_E t/\hbar}\mathcal{U}(t,0).
\label{eq:rhotildeS+E}
\end{align}
Taking the time derivative of Eq. \ref{eq:rhotildeS+E} gives:
\begin{align}
\dot{\tilde{\rho}}_{S+E} =
 \frac{\partial \mathcal{U}(0,t)}{\partial t} e^{i H_E t/\hbar} \rho_{S+E} e^{-i H_E t/\hbar} \, \mathcal{U}(t,0) +
  \mathcal{U}(0,t) e^{i H_E t/\hbar} \rho_{S+E} e^{-i H_E t/\hbar}\frac{\partial \mathcal{U}(t,0) }{\partial t} \nonumber \\
   +   \mathcal{U}(0,t) \frac{\partial}{\partial t} \left( e^{i H_E t/\hbar} \rho_{S+E} e^{-i H_E t/\hbar}\right)\mathcal{U}(t,0) 
\end{align}
Applying the Eqs. \ref{eq:HU} and \ref{eq:UH} for the time derivative for $\mathcal U$ and $\mathcal U^\dagger $, and  using Eq.\ref{eq:vNAppB} for the time derivative of $\rho_{S+E}$, and using the fact that the system and environment Hamiltonians operate in different Hilbert spaces and thus commute with each other (but not with the interaction Hamiltonian):
\begin{align}
\dot{\tilde{\rho}}_{S+E} =
 \mathcal{U}(0,t) \left( [\frac{i}{\hbar}H_S, e^{i H_E t/\hbar} \rho_{S+E} e^{-i H_E t/\hbar}]  +e^{i H_E t/\hbar}[\frac{i}{\hbar}H_E,  \rho_{S+E}] e^{-i H_E t/\hbar} + e^{i H_E t/\hbar}  [\frac{-i}{\hbar}H, \rho_{S+E}] e^{-i H_E t/\hbar} \right) \, \mathcal{U}(t,0) 
 \nonumber \\
 =\frac{-i}{\hbar} \,\mathcal{U}(0,t)  e^{i H_E t/\hbar} [H_I,  \rho_{S+E}] e^{-i H_E t/\hbar} \, \mathcal{U}(t,0) .
\end{align}
The von Neumann equation for the interaction picture density matrix of the combined dot plus the lead, $\tilde{\rho}_{S+E}$, becomes 
\begin{align}
\dot{\tilde{\rho}}_{S+E} = \frac{-i}{\hbar}[\tilde{H}_I,\tilde\rho_{S+E}]
\label{eq:vN1App}
\end{align}
with 
\begin{align}
\tilde{H}_I = \mathcal{U}(0,t)  e^{i H_E t/\hbar} H_I  e^{-i H_E t/\hbar} \, \mathcal{U}(t,0).
\end{align}

\section{Thermal environmental density matrix implies first-order term is zero}
 \label{App:1stTerm0} 
 Equation \ref{eq:vN2ndO} is:
 \begin{align}
\dot{\tilde{\rho}}_{S+E}(t_1) =&\frac{-i}{\hbar} [\tilde{H}_I(t_1),\tilde{\rho}_{S+E}(t_0)] 
+\frac{(-i)^2}{\hbar^2} \int_{t_0}^{t_1} dt_2 [\tilde{H}_I(t_1) ,[\tilde{H}_I(t_2),\tilde{\rho}_{S+E}(t_2)]]
\label{eq:vN2ndOAppC}
\end{align}
 where the interaction Hamiltonian consists of sums of pairs of operators in which an electron is created in the dot and destroyed in the lead or vice versa:
 \begin{equation}
H_{I}= \sum_l \lambda_{l,\uparrow} \hat a^\dagger_{l,\uparrow} \hat a_\uparrow + \lambda_{l,\downarrow} \hat a ^\dagger_{l,\downarrow} \hat a_\downarrow +\lambda_{l,\uparrow} \hat a^\dagger_\uparrow \hat a_{l,\uparrow} + \lambda_{l,\uparrow} \hat a ^\dagger_\downarrow \hat a_{l,\downarrow}.
\end{equation}
We wish to show that when we take the trace of  Eq. \ref{eq:vN2ndOAppC} over the environment, the first term is zero.  
 
In order to describe an ordinary metal in which screening mitigates electron correlation, we use the independent electron approximation for the lead.  This allows us to consider the lead's state as a thermal mixture of Slater determinants of non-interacting electrons in equilibrium at constant temperature and chemical potential.  Each Slater determinant state, $|\Phi_i^M\rangle$ is an eigenstate of $\hat{H}_E - \mu \hat{M}$ (and of $\hat{H}_E $) :
\begin{align}
\hat{H}_E - \mu \hat{M} |\phi_i^M\rangle = E_i^M |\phi_i^M\rangle  \\
\hat{H}_E |\phi_i^M\rangle = \mathcal{E}_i^M |\phi_i^M\rangle 
\end{align}
with eigenvalues that depend on how many electrons there are ($M$) and which single-particle states are filled (where $i$ is a single value that enumerates each possible many-body combination of single-particle states $k \equiv (l,s)$ that are occupied by the $M$ electrons and $n_k^{i,M}$ is the occupancy of that $k$th state in the $i$th combination, either 0 or 1). Much of the argument below is from the textbook \cite{Mattuck1992}.

The partition function for this lead is then:
\begin{align}
Z_0 = \sum_{i,M} e^{-\beta E_i^M}= \sum_{i,M} \prod_{k=1}^M  e^{-\beta ( \varepsilon_{k_1} - \mu)n_k^{i,M}}
=  \prod_{k=1}^M 1+e^{-\beta ( \varepsilon_{k} - \mu)},
\label{eq:PartFxn}
\end{align}
($\beta \equiv 1/k_{\rm B} T$)
and the density matrix for the lead is:
\begin{align}
\rho_E = \frac{e^{-\beta (\hat{H}_E - \mu\hat{M})}}{Z_0}=\frac{1}{Z_0}\sum_{i,M} e^{-\beta E_i^M} |\phi_i^M\rangle \langle \phi_i^M|.
\end{align}
This density matrix is diagonal in eigenstates. 
In particular, there are no coherences between states of different particle number.
The trace of the first term in Eq. \ref{eq:vN2ndOAppC} is  proportional to:
\begin{align}
 {\rm Tr}_E\{ [\tilde{H}_I (t_1), \tilde{\rho}_{S+E}(t_0)] \} \equiv \sum_{i,M}\langle \phi_i^M| \, [\tilde{H}_I (t_1), \tilde{\rho}_{S+E}(t_0)] |\phi_i^M\rangle
\label{eq:TraceEnv2}
\end{align}
Examining a single term of $\tilde{H}_I$, and assuming that the system and lead density matrices are separable:
\begin{align}
 {\rm Tr}_E\{ [\tilde{a}_{l,s}^\dagger(t_1) \tilde{a}_s(t_1), \tilde{\rho}_{S+E}(t_0)] \} = \sum_{i,M}\langle \phi_i^M| \, [\tilde{a}_{l,s}^\dagger(t_1) \tilde{a}_s(t_1), \tilde{\rho}_{S}(t_0)\otimes \tilde{\rho}_{E}(t_0)] |\phi_i^M\rangle.
\label{eq:TraceEnv1Term}
\end{align}
The system operators act only on the system density matrix and commute with all the environment operators and the environment density matrix.  Similarly the environment operators act only on the environment density matrix, so:
\begin{align}
 {\rm Tr}_E\{ [\tilde{a}_{l,s}^\dagger(t_1) \tilde{a}_s(t_1), \tilde{\rho}_{S+E}(t_0)] \} = \sum_{i,M}\langle \phi_i^M| \, [\tilde{a}_{l,s}^\dagger(t_1),  \tilde{\rho}_{E}(t_0)] 
 |\phi_i^M\rangle
 \otimes [ \tilde{a}_s(t_1),\tilde{\rho}_{S}(t_0)]=0.
\label{eq:TraceEnv1TermSep}
\end{align}
Because the environment's Hamiltonian is number conserving, transforming the creation operator into the interaction picture does not change the fact that it creates one particle, and transforming the environment density matrix into the interaction picture does not change that it is diagonal in states with definite particle number.

\section{Correlation functions of the lead}
\label{App:ThermalLead}
%
%
In this  appendix we omit factors of $\hbar$ and instead indicate by text where they should appear.  All  energies in  the complex exponentials below should be divided by $\hbar$. All energies in exponentials with $\beta = 1/k_{\rm B}T$ are actually energies.

The second order term of Eq. \ref{eq:vN2ndOAppC} is not zero when we trace over the environment.  It leads instead to terms such as:
\begin{align}
N_{ls}(t_1, t_2) \equiv {\rm Tr}_B \left\{ \tilde{a}_{l,s}^\dagger (t_1)  \tilde{a}_{l,s} (t_2)    \rho_{B}(0)  \right\}
= {\rm Tr}_B \left\{ \frac{e^{-\beta (\hat{H}_E - \mu\hat{M})}}{Z_0}  e^{i\hat{H}_E t_1} a_{l,s}^\dagger  e^{-i\hat{H}_E t_1} e^{i\hat{H}_E t_2} \tilde{a}_{l,s} e^{-i\hat{H}_E t_2}  \right\}
\nonumber \\
= \sum_{i,M} \frac{e^{-\beta E_i^M}}{Z_0} \langle \phi_i^M|  e^{i\hat{H}_E t_1} a_{l,s}^\dagger  e^{-i\hat{H}_E (t_1-t_2)} a_{l,s} e^{-i\hat{H}_E t_2} |\phi_i^M\rangle
\nonumber \\
= \sum_{i,M} \frac{e^{-\beta E_i^M}}{Z_0} e^{i \mathcal{E}_i^M  t_1} e^{-i \mathcal{E}_i^M t_2}\langle \phi_i^M|   a_{l,s}^\dagger  e^{-i\hat{H}_E (t_1 - t_2)}a_{l,s} |\phi_i^M\rangle
= \sum_{i,M} \frac{e^{-\beta E_i^M}}{Z_0} e^{i \mathcal{E}_i^M ( t_1-t_2)} e^{-i (\mathcal{E}_i^M   - \varepsilon_{ls})(t_1 -t_2)}\langle \phi_i^M|   a_{l,s}^\dagger a_{l,s}|\phi_i^M\rangle
\nonumber \\
= e^{i \varepsilon_{ls}(t_1 -t_2)} \sum_{i,M} \frac{e^{-\beta E_i^M}}{Z_0} \langle \phi_i^M|   n_{l,s}  |\phi_i^M\rangle.
\end{align}

This can be further evaluated similarly to the expression for the partition function, Eq. \ref{eq:PartFxn}, recognizing that in the Fermi liquid approximation the MB states of the lead are determinants of states $k$ which are occupied or unoccupied:
\begin{align}
N_{ls}(t_1, t_2) 
= e^{i \varepsilon_{ls}(t_1 -t_2)} \sum_{i,M} \frac{e^{-\beta E_i^M}}{Z_0} \langle \phi_i^M|   n_{l,s}  |\phi_i^M\rangle
=  e^{i \varepsilon_{ls}(t_1 -t_2)} \sum_{i,M} \prod_{k=1}^{M_{sites}} \frac{e^{-\beta (\varepsilon_k - \mu)n_k^{i,M}}}{Z_0} \langle \phi_i^M|   n_{l,s}  |\phi_i^M\rangle
\nonumber \\
=  e^{i \varepsilon_{ls}(t_1 -t_2)} \sum_{i,M} \prod_{k=1}^{M_{sites}} \frac{e^{-\beta (\varepsilon_k - \mu)n_k^{i,M}}}{1+ e^{-\beta ( \varepsilon_{k} - \mu)}} \langle \phi_i^M|   n_{l,s}  |\phi_i^M\rangle
\end{align}
We separate the product explicitly into the cases when $k=l,s$ and when it does not:
\begin{align}
N_{ls}(t_1, t_2) 
=   e^{i \varepsilon_{ls}(t_1 -t_2)} \sum_{i,M} \langle \phi_i^M|   n_{l,s}  |\phi_i^M\rangle  \frac{e^{-\beta (\varepsilon_{l,s} - \mu)n_{ls}^{i,M}}}{1+ e^{-\beta ( \varepsilon_{l,s} - \mu)}} \prod_{k\neq (l,s)}^{M_{sites}} \frac{e^{-\beta (\varepsilon_k - \mu)n_k^{i,M}}}{1+ e^{-\beta ( \varepsilon_{k} - \mu)}}
\end{align}
The numerator of the product with $k\neq l,s$ is similar to the partition function calculations, except with one particular state omitted.  This leaves the numerator and denominator identical and the product over $k\neq l,s$ reduces to one, leaving:
\begin{align}
N_{ls}(t_1, t_2) 
=   e^{i \varepsilon_{ls}(t_1 -t_2)} \sum_{i,M} \langle \phi_i^M|   n_{l,s}  |\phi_i^M\rangle  \frac{e^{-\beta (\varepsilon_{l,s} - \mu)n_{ls}^{i,M}}}{1+ e^{-\beta ( \varepsilon_{l,s} - \mu)}}
=   e^{i \varepsilon_{ls}(t_1 -t_2)}  \frac{0e^0 + 1e^{-\beta (\varepsilon_{l,s} - \mu)}}{1+ e^{-\beta ( \varepsilon_{l,s} - \mu)}}
\nonumber \\
=   e^{i \varepsilon_{ls}(t_1 -t_2)}  \frac{1}{1+ e^{\beta ( \varepsilon_{l,s} - \mu)}}
=   e^{i \varepsilon_{ls}(t_1 -t_2)}  n_e(\varepsilon_{l,s}, \mu, T)
\end{align}
Similarly,
\begin{align}
G_{ls}(t_1, t_2) &\equiv {\rm Tr}_B \left\{ \tilde{a}_{l,s} (t_1)  \tilde{a}_{l,s}^\dagger (t_2)    \rho_{B}(0)  \right\}
= {\rm Tr}_B \left\{ \frac{e^{-\beta (\hat{H}_E - \mu\hat{M})}}{Z_0}  e^{i\hat{H}_E t_1} a_{l,s}  e^{-i\hat{H}_E t_1} e^{i\hat{H}_E t_2} \tilde{a}_{l,s}^\dagger e^{-i\hat{H}_E t_2}  \right\}
\nonumber \\
&= \sum_{i,M} \frac{e^{-\beta E_i^M}}{Z_0} \langle \phi_i^M|  e^{i\hat{H}_E t_1} a_{l,s}  e^{-i\hat{H}_E (t_1-t_2)} a_{l,s}^\dagger e^{-i\hat{H}_E t_2} |\phi_i^M\rangle
\nonumber \\
&= \sum_{i,M} \frac{e^{-\beta E_i^M}}{Z_0} e^{i \mathcal{E}_i^M  t_1} e^{-i \mathcal{E}_i^M t_2}\langle \phi_i^M|   a_{l,s}  e^{-i\hat{H}_E (t_1 - t_2)}a_{l,s}^\dagger |\phi_i^M\rangle
\nonumber \\
&= \sum_{i,M} \frac{e^{-\beta E_i^M}}{Z_0} e^{i \mathcal{E}_i^M ( t_1-t_2)} e^{-i (\mathcal{E}_i^M   + \varepsilon_{ls})(t_1 -t_2)}\langle \phi_i^M|   a_{l,s} a_{l,s}^\dagger |\phi_i^M\rangle
\nonumber \\
&= \sum_{i,M} \frac{e^{-\beta E_i^M}}{Z_0}  e^{-i( \varepsilon_{ls} )(t_1 -t_2)}\langle \phi_i^M|   a_{l,s} a_{l,s}^\dagger |\phi_i^M\rangle
=e^{-i\varepsilon_{ls} (t_1 -t_2)} \sum_{i,M} \frac{e^{-\beta E_i^M}}{Z_0}  \langle \phi_i^M|   1-n_{l,s} |\phi_i^M\rangle
\nonumber \\
&= e^{-i \varepsilon_{ls}(t_1 -t_2)} (1- n_e(\varepsilon_{l,s}, \mu, T)) \equiv  e^{-i \varepsilon_{ls}(t_1 -t_2)}  n_h(\varepsilon_{l,s}, \mu, T).
\end{align}
Summarizing (recall that $t_1\geq t_2)$:
\begin{align}
G_{ls}(t_1, t_2) 
=e^{-i \varepsilon_{ls}(t_1 -t_2)} \frac{1}{1+ e^{-\beta ( \varepsilon_{l,s} - \mu)}} =e^{-i \varepsilon_{ls}(t_1 -t_2)}  n_h(\varepsilon_{l,s}, \mu, T).
\\
N_{ls}(t_1, t_2) 
=   e^{i \varepsilon_{ls}(t_1 -t_2)}  \frac{1}{1+ e^{\beta ( \varepsilon_{l,s} - \mu)}}
=   e^{i \varepsilon_{ls}(t_1 -t_2)}  n_e(\varepsilon_{l,s}, \mu, T)
\end{align}
These correlation functions are unitless.

We'll need to evaluate the Fourier transforms from $t_1 - t_2$ to frequency space 
 of these time correlation functions, and a sum over lead states, $l$, of those frequency space correlation functions.  We'll transform that sum to an integral using a density of states for the lead, $D(\varepsilon)$, restoring the factors of $\hbar$:
\begin{align}
&N_{ls} (\omega) \equiv \int_{-\infty}^{\infty} \mathrm{d}u  \, N_{ls}(u) e^{-i\omega u}  
= \int_{-\infty}^{\infty} \mathrm{d}u  \, e^{i \varepsilon_{ls}u/\hbar}  n_e(\varepsilon_{l,s}, \mu, T) e^{-i\omega u}  =  n_e(\varepsilon_{l,s}, \mu, T)2\pi \delta(\omega - \varepsilon_{l,s}/\hbar) 
 \\
&\sum_l N_{ls} (\omega)  = \int_{-\infty}^{\infty} \mathrm{d}\varepsilon D(\varepsilon) n_e(\varepsilon, \mu, T)2\pi\hbar \delta(	\hbar \omega - \varepsilon) 
= D(\hbar\omega) n_e(\hbar\omega, \mu, T) 2\pi \hbar
\end{align}
which used $\delta(x/c) = c \delta(x)$.
\begin{align}
&G_{ls} (\omega) \equiv \int_{-\infty}^{\infty} \mathrm{d}u  \, G_{ls}(u) e^{-i\omega u}   
= \int_{-\infty}^{\infty} \mathrm{d}u  \, e^{-i \varepsilon_{ls}u/\hbar}  n_h(\varepsilon_{l,s}, \mu, T) e^{-i\omega u}   =  n_h(\varepsilon_{l,s}, \mu, T) 2\pi \delta(\omega +\varepsilon_{l,s}/\hbar)
\\
&\sum_l G_{ls} (\omega)= \int_{-\infty}^{\infty} \mathrm{d}\varepsilon D(\varepsilon) n_h(\varepsilon, \mu, T)2\pi \hbar \delta(\omega + \varepsilon) 
= D(-\hbar \omega) n_h(-\hbar \omega, \mu, T) 2\pi \hbar
\end{align}

Our correlation functions for a particular energy level with quantum numbers $l,s$ can be summed over all energies to determine the correlation time for the environment:
\begin{align}
C(t) \equiv \langle \tilde{a}(t) \tilde{a}^\dagger(0) \rangle = \int_{-\infty}^{\infty} d \epsilon \frac{e^{-i\epsilon t/\hbar} }{1+e^{\beta(\epsilon - \mu)}}
\end{align}
This integral has poles when $e^{\beta(\epsilon - \mu)} = -1$, i.e. when $\epsilon = \mu \pm i\frac{\pi}{\beta}(2n+1)$.  By choosing the causal contour the sum over residues gives:
\begin{align}
C(t) = \frac{-2\pi i}{\beta} \sum_{n=0}^{\infty} e^{-i\mu t/\hbar}  e^{-t\pi (2n+1)/\beta \hbar } =  \frac{-2\pi i}{\beta}  e^{-i\mu t/\hbar} \frac{ e^{-\pi t/\beta\hbar}}{1 -  e^{-2\pi  t/\beta \hbar}} \sim e^{-it/ \tau_c }
\end{align}
where $\tau_c \equiv \beta \hbar/ \pi =\frac{\hbar}{\pi k_{\rm B} T } $.  Additional degrees of freedom in the lead and/or the finite band width have the potential to further reduce the correlation time.

 \section{Results of matrix multiplication before secular approximation}
 \label{App:MatrixMult}
 
In Matlab we denote $ \mathcal{A}(t-t_0)$ as $\mathrm{A_{t t_o}}$, $ e^{-iE_b(0-t)} \rightarrow \mathrm{Coul_{0t}}$ etc., and the complex conjugate $ B^*(t,t_0)\rightarrow \overline{\mathrm{B_{t t_0}}}$ (which we note does NOT equal $\mathcal{B}(t_0,t)$).  In addition, we've replaced $\tau \rightarrow u$, $t_1 \rightarrow t$, and $t_1 - \tau \rightarrow d = t-u$,  and $\lambda^2/\hbar^2 \rightarrow 1$. The bath correlation functions do have the property that $G_{ls}(-\tau) = G_{ls}^*(\tau) $, or $\overline{\mathrm{G_{-u}^\uparrow}} = \mathrm{G_{u}^\uparrow}$. 

The first diagonal element yields:
\begin{align}
  -\frac{d\rho_{0,0}}{dt} = \frac{\lambda^2}{\hbar^2} \sum \int du
  &\left(\mathrm{A_{0t}}\,\overline{\mathrm{A_{0d}}}\mathrm{N_{-u}^\uparrow}\,+\mathrm{B_{0t}}\,\overline{\mathrm{B_{0d}}}\mathrm{N_{-u}^\downarrow}\,+\mathrm{C_{0t}}\,\overline{\mathrm{C_{0d}}}\,\mathrm{N_{-u}^\uparrow} +\mathrm{D_{0t}}\,\overline{\mathrm{D_{0d}}}\,\mathrm{N_{-u}^\downarrow} 
  \right. \nonumber \\ 
   &  \left. \qquad + \mathrm{A_{0d}}\, \overline{\mathrm{A_{0t}}}\,\overline{\mathrm{N_{-u}^\uparrow}}+\mathrm{B_{0d}}\,\overline{\mathrm{B_{0t}}}\,\overline{\mathrm{N_{-u}^\downarrow}}+\mathrm{C_{0d}}\,\overline{\mathrm{C_{0t}}}\,\overline{\mathrm{N_{-u}^\uparrow}}+\mathrm{D_{0d}}\,\overline{\mathrm{D_{0t}}}\,\overline{\mathrm{N_{-u}^\downarrow}}\right)\,\tilde{\rho}_{00}
  \nonumber \\
  +&\left(-\mathrm{A_{0d}}\,\overline{\mathrm{A_{0t}}}\,\mathrm{G_{-u}^\uparrow}-\mathrm{B_{0d}}\,\overline{\mathrm{B_{0t}}}\,\mathrm{G_{-u}^\downarrow}-\mathrm{A_{0t}}\,\overline{\mathrm{A_{0d}}}\,\overline{\mathrm{G_{-u}^\uparrow}}-\mathrm{B_{0t}}\,\overline{\mathrm{B_{0d}}}\,\overline{\mathrm{G_{-u}^\downarrow}}\right)\,\mathrm{\tilde{\rho}_{\uparrow \uparrow}}
  \nonumber \\
  +&\left(-\mathrm{C_{0d}}\,\overline{\mathrm{A_{0t}}}\,\mathrm{G_{-u}^\uparrow}-\mathrm{D_{0d}}\,\overline{\mathrm{B_{0t}}}\,\mathrm{G_{-u}^\downarrow} - \mathrm{C_{0t}}\,\overline{\mathrm{A_{0d}}}\,\overline{\mathrm{G_{-u}^\uparrow}}-\mathrm{D_{0t}}\,\overline{\mathrm{B_{0d}}}\,\overline{\mathrm{G_{-u}^\downarrow}}\right)\,\mathrm{\tilde{\rho}_{\uparrow \downarrow}}
  \nonumber \\
  +&\left(-\mathrm{A_{0d}}\,\overline{\mathrm{C_{0t}}}\,\mathrm{G_{-u}^\uparrow}-\mathrm{B_{0d}}\,\overline{\mathrm{D_{0t}}}\,\mathrm{G_{-u}^\downarrow} - \mathrm{A_{0t}}\,\overline{\mathrm{C_{0d}}}\,\overline{\mathrm{G_{-u}^\uparrow}}-\mathrm{B_{0t}}\,\overline{\mathrm{D_{0d}}}\,\overline{\mathrm{G_{-u}^\downarrow}}\right)\,\mathrm{\tilde{\rho}_{\downarrow \uparrow}}
  \nonumber \\
  +&\left(-\mathrm{C_{0d}}\,\overline{\mathrm{C_{0t}}}\,\mathrm{G_{-u}^\uparrow} - \mathrm{D_{0d}}\,\overline{\mathrm{D_{0t}}}\,\mathrm{G_{-u}^\downarrow}-\mathrm{C_{0t}}\,\overline{\mathrm{C_{0d}}}\,\overline{\mathrm{G_{-u}^\uparrow}}-\mathrm{D_{0t}}\,\overline{\mathrm{D_{0d}}}\,\overline{\mathrm{G_{-u}^\downarrow}}\right)\,\mathrm{\tilde{\rho}_{\downarrow \downarrow}}
\end{align}

The second diagonal element yields:
\begin{align}
  -\frac{d\rho_{\uparrow \uparrow}}{dt} =   \frac{\lambda^2}{\hbar^2} \sum \int du &\left(- \mathrm{A_{0t}}\,\overline{\mathrm{A_{0d}}}\mathrm{N_{-u}^\uparrow}-\mathrm{B_{0t}}\,\overline{\mathrm{B_{0d}}}\mathrm{N_{-u}^\downarrow} - \mathrm{A_{0d}}\,\overline{\mathrm{A_{0t}}}\,\overline{\mathrm{N_{-u}^\uparrow}}-\mathrm{B_{0d}}\,\overline{\mathrm{B_{0t}}}\,\overline{\mathrm{N_{-u}^\downarrow}}\right)\,\tilde{\rho}_{00}
  \nonumber \\
  &  +\left(\mathrm{A_{0d}}\,\overline{\mathrm{A_{0t}}}\mathrm{G_{-u}^\uparrow}\,+\mathrm{B_{0d}}\,\overline{\mathrm{B_{0t}}}\mathrm{G_{-u}^\downarrow}\,+\mathrm{A_{0t}}\,\overline{\mathrm{A_{0d}}}\,\overline{\mathrm{G_{-u}^\uparrow}}+\mathrm{B_{0t}}\,\overline{\mathrm{B_{0d}}}\,\overline{\mathrm{G_{-u}^\downarrow}}
 \right. \nonumber \\
  & \left. \qquad +\mathrm{A_{0d}}\,\mathrm{Coul_{0t}}\,\overline{\mathrm{A_{0t}}}\,\overline{\mathrm{Coul_{0d}}}\mathrm{N_{-u}^\downarrow}\,+\mathrm{B_{0d}}\,\mathrm{Coul_{0t}}\,\overline{\mathrm{B_{0t}}}\,\overline{\mathrm{Coul_{0d}}}\mathrm{N_{-u}^\uparrow}\,
   \right. \nonumber \\
  & \left. \qquad
  +\mathrm{A_{0t}}\,\mathrm{Coul_{0d}}\,\overline{\mathrm{A_{0d}}}\,\overline{\mathrm{Coul_{0t}}}\,\overline{\mathrm{N_{-u}^\downarrow}}+\mathrm{B_{0t}}\,\mathrm{Coul_{0d}}\,\overline{\mathrm{B_{0d}}}\,\overline{\mathrm{Coul_{0t}}}\,\overline{\mathrm{N_{-u}^\uparrow}}\right)\,\mathrm{\tilde{\rho}_{\uparrow \uparrow}}
  \nonumber \\
 & +\left(\mathrm{C_{0d}}\,\overline{\mathrm{A_{0t}}}\mathrm{G_{-u}^\uparrow}\,+\mathrm{D_{0d}}\,\overline{\mathrm{B_{0t}}}\mathrm{G_{-u}^\downarrow}\,+\mathrm{C_{0d}}\,\mathrm{Coul_{0t}}\,\overline{\mathrm{A_{0t}}}\,\overline{\mathrm{Coul_{0d}}}\mathrm{N_{-u}^\downarrow}\,+\mathrm{Coul_{0t}}\,\mathrm{D_{0d}}\,\overline{\mathrm{B_{0t}}}\,\overline{\mathrm{Coul_{0d}}}\mathrm{N_{-u}^\uparrow}\,\right)\,\mathrm{\tilde{\rho}_{\uparrow \downarrow}}
  \nonumber \\
  &+\left(\mathrm{A_{0t}}\,\overline{\mathrm{C_{0d}}}\,\overline{\mathrm{G_{-u}^\uparrow}}+\mathrm{B_{0t}}\,\overline{\mathrm{D_{0d}}}\,\overline{\mathrm{G_{-u}^\downarrow}}+\mathrm{A_{0t}}\,\mathrm{Coul_{0d}}\,\overline{\mathrm{C_{0d}}}\,\overline{\mathrm{Coul_{0t}}}\,\overline{\mathrm{N_{-u}^\downarrow}}+\mathrm{B_{0t}}\,\mathrm{Coul_{0d}}\,\overline{\mathrm{Coul_{0t}}}\,\overline{\mathrm{D_{0d}}}\,\overline{\mathrm{N_{-u}^\uparrow}}\right)\,\mathrm{\tilde{\rho}_{\downarrow \uparrow}}
\nonumber \\
  &+\left(-\mathrm{A_{0t}}\,\mathrm{Coul_{0d}}\,\overline{\mathrm{A_{0d}}}\,\overline{\mathrm{Coul_{0t}}}\mathrm{G_{-u}^\downarrow}\,-\mathrm{B_{0t}}\,\mathrm{Coul_{0d}}\,\overline{\mathrm{B_{0d}}}\,\overline{\mathrm{Coul_{0t}}}\mathrm{G_{-u}^\uparrow}\,
   \right. \nonumber \\
  & \left. \qquad
  -\mathrm{A_{0d}}\,\mathrm{Coul_{0t}}\,\overline{\mathrm{A_{0t}}}\,\overline{\mathrm{Coul_{0d}}}\,\overline{\mathrm{G_{-u}^\downarrow}}-\mathrm{B_{0d}}\,\mathrm{Coul_{0t}}\,\overline{\mathrm{B_{0t}}}\,\overline{\mathrm{Coul_{0d}}}\,\overline{\mathrm{G_{-u}^\uparrow}}\right)\,\mathrm{\tilde{\rho}_{bb}}
  \end{align}

The third diagonal element yields:
\begin{align}
  -\frac{d\rho_{\downarrow \downarrow}}{dt} =  \frac{\lambda^2}{\hbar^2}\sum \int du
  &\left(-\mathrm{C_{0t}}\,\mathrm{N_{-u}^\uparrow}\,\overline{\mathrm{C_{0d}}}-\mathrm{D_{0t}}\,\mathrm{N_{-u}^\downarrow}\,\overline{\mathrm{D_{0d}}}-\mathrm{C_{0d}}\,\overline{\mathrm{C_{0t}}}\,\overline{\mathrm{N_{-u}^\uparrow}}-\mathrm{D_{0d}}\,\overline{\mathrm{D_{0t}}}\,\overline{\mathrm{N_{-u}^\downarrow}}\right)\,\tilde{\rho}_{00}
  \nonumber \\
  +&\left(\mathrm{C_{0t}}\,\overline{\mathrm{A_{0d}}}\,\overline{\mathrm{G_{-u}^\uparrow}}+\mathrm{D_{0t}}\,\overline{\mathrm{B_{0d}}}\,\overline{\mathrm{G_{-u}^\downarrow}}+\mathrm{C_{0t}}\,\mathrm{Coul_{0d}}\,\overline{\mathrm{A_{0d}}}\,\overline{\mathrm{Coul_{0t}}}\,\overline{\mathrm{N_{-u}^\downarrow}}+\mathrm{Coul_{0d}}\,\mathrm{D_{0t}}\,\overline{\mathrm{B_{0d}}}\,\overline{\mathrm{Coul_{0t}}}\,\overline{\mathrm{N_{-u}^\uparrow}}\right)\,\mathrm{\tilde{\rho}_{\uparrow \downarrow}}
  \nonumber \\
  +&\left(\mathrm{A_{0d}}\,\mathrm{G_{-u}^\uparrow}\,\overline{\mathrm{C_{0t}}}+\mathrm{B_{0d}}\,\mathrm{G_{-u}^\downarrow}\,\overline{\mathrm{D_{0t}}}+\mathrm{A_{0d}}\,\mathrm{Coul_{0t}}\,\mathrm{N_{-u}^\downarrow}\,\overline{\mathrm{C_{0t}}}\,\overline{\mathrm{Coul_{0d}}}+\mathrm{B_{0d}}\,\mathrm{Coul_{0t}}\,\mathrm{N_{-u}^\uparrow}\,\overline{\mathrm{Coul_{0d}}}\,\overline{\mathrm{D_{0t}}}\right)\,\mathrm{\tilde{\rho}_{\downarrow \uparrow}}
  \nonumber \\
  +&\left(\mathrm{C_{0d}}\,\mathrm{G_{-u}^\uparrow}\,\overline{\mathrm{C_{0t}}}+\mathrm{D_{0d}}\,\mathrm{G_{-u}^\downarrow}\,\overline{\mathrm{D_{0t}}}+\mathrm{C_{0t}}\,\overline{\mathrm{C_{0d}}}\,\overline{\mathrm{G_{-u}^\uparrow}}+\mathrm{D_{0t}}\,\overline{\mathrm{D_{0d}}}\,\overline{\mathrm{G_{-u}^\downarrow}}
 +\mathrm{C_{0d}}\,\mathrm{Coul_{0t}}\,\mathrm{N_{-u}^\downarrow}\,\overline{\mathrm{C_{0t}}}\,\overline{\mathrm{Coul_{0d}}}
 \right.  \nonumber \\
  &\hspace{1.5em} \left.
  +\mathrm{Coul_{0t}}\,\mathrm{D_{0d}}\,\mathrm{N_{-u}^\uparrow}\,\overline{\mathrm{Coul_{0d}}}\,\overline{\mathrm{D_{0t}}}
  +\mathrm{C_{0t}}\,\mathrm{Coul_{0d}}\,\overline{\mathrm{C_{0d}}}\,\overline{\mathrm{Coul_{0t}}}\,\overline{\mathrm{N_{-u}^\downarrow}}+\mathrm{Coul_{0d}}\,\mathrm{D_{0t}}\,\overline{\mathrm{Coul_{0t}}}\,\overline{\mathrm{D_{0d}}}\,\overline{\mathrm{N_{-u}^\uparrow}}\right)\,\mathrm{\tilde{\rho}_{\downarrow \downarrow}}
  \nonumber \\
  +&\left(-\mathrm{C_{0t}}\,\mathrm{Coul_{0d}}\,\mathrm{G_{-u}^\downarrow}\,\overline{\mathrm{C_{0d}}}\,\overline{\mathrm{Coul_{0t}}}
  -\mathrm{Coul_{0d}}\,\mathrm{D_{0t}}\,\mathrm{G_{-u}^\uparrow}\,\overline{\mathrm{Coul_{0t}}}\,\overline{\mathrm{D_{0d}}}
  -\mathrm{C_{0d}}\,\mathrm{Coul_{0t}}\,\overline{\mathrm{C_{0t}}}\,\overline{\mathrm{Coul_{0d}}}\,\overline{\mathrm{G_{-u}^\downarrow}}
  \right.  \nonumber \\
  &\qquad \left.
  -\mathrm{Coul_{0t}}\,\mathrm{D_{0d}}\,\overline{\mathrm{Coul_{0d}}}\,\overline{\mathrm{D_{0t}}}\,\overline{\mathrm{G_{-u}^\uparrow}}\right)\,\mathrm{\tilde{\rho}_{bb}}
\end{align}

The 4th diagonal element yields:
\begin{align}
  -\frac{d\rho_{b,b}}{dt} =  \frac{\lambda^2}{\hbar^2} \sum \int du
  &\left(-\mathrm{A_{0d}}\,\mathrm{Coul_{0t}}\,\overline{\mathrm{A_{0t}}}\,\overline{\mathrm{Coul_{0d}}}\,\mathrm{N_{-u}^\downarrow} - \mathrm{B_{0d}}\,\mathrm{Coul_{0t}}\,\overline{\mathrm{B_{0t}}}\,\overline{\mathrm{Coul_{0d}}}\,\mathrm{N_{-u}^\uparrow} 
  - \mathrm{A_{0t}}\,\mathrm{Coul_{0d}}\,\overline{\mathrm{A_{0d}}}\,\overline{\mathrm{Coul_{0t}}}\,\overline{\mathrm{N_{-u}^\downarrow}}
  \right.  \nonumber \\
  &\qquad \left.
  -\mathrm{B_{0t}}\,\mathrm{Coul_{0d}}\,\overline{\mathrm{B_{0d}}}\,\overline{\mathrm{Coul_{0t}}}\,\overline{\mathrm{N_{-u}^\uparrow}}\right)\,\mathrm{\tilde{\rho}_{\uparrow \uparrow}}
  \nonumber \\
  +&\left(-\mathrm{C_{0d}}\,\mathrm{Coul_{0t}}\,\overline{\mathrm{A_{0t}}}\,\overline{\mathrm{Coul_{0d}}}\,\mathrm{N_{-u}^\downarrow}
   - \mathrm{Coul_{0t}}\,\mathrm{D_{0d}}\,\overline{\mathrm{B_{0t}}}\,\overline{\mathrm{Coul_{0d}}}\,\mathrm{N_{-u}^\uparrow} 
   - \mathrm{C_{0t}}\,\mathrm{Coul_{0d}}\,\overline{\mathrm{A_{0d}}}\,\overline{\mathrm{Coul_{0t}}}\,\overline{\mathrm{N_{-u}^\downarrow}}
   \right.  \nonumber \\
  &\qquad \left.
  -\mathrm{Coul_{0d}}\,\mathrm{D_{0t}}\,\overline{\mathrm{B_{0d}}}\,\overline{\mathrm{Coul_{0t}}}\,\overline{\mathrm{N_{-u}^\uparrow}}\right)\,\mathrm{\tilde{\rho}_{\uparrow \downarrow}}
  \nonumber \\
  +&\left(-\mathrm{A_{0d}}\,\mathrm{Coul_{0t}}\,\overline{\mathrm{C_{0t}}}\,\overline{\mathrm{Coul_{0d}}}\,\mathrm{N_{-u}^\downarrow} 
  - \mathrm{B_{0d}}\,\mathrm{Coul_{0t}}\,\overline{\mathrm{Coul_{0d}}}\,\overline{\mathrm{D_{0t}}}\,\mathrm{N_{-u}^\uparrow}
   - \mathrm{A_{0t}}\,\mathrm{Coul_{0d}}\,\overline{\mathrm{C_{0d}}}\,\overline{\mathrm{Coul_{0t}}}\,\overline{\mathrm{N_{-u}^\downarrow}}
   \right.  \nonumber \\
  &\qquad \left.
  -\mathrm{B_{0t}}\,\mathrm{Coul_{0d}}\,\overline{\mathrm{Coul_{0t}}}\,\overline{\mathrm{D_{0d}}}\,\overline{\mathrm{N_{-u}^\uparrow}}\right)\,\mathrm{\tilde{\rho}_{\downarrow \uparrow}}
  \nonumber \\
  +&\left(-\mathrm{C_{0d}}\,\mathrm{Coul_{0t}}\,\overline{\mathrm{C_{0t}}}\,\overline{\mathrm{Coul_{0d}}}\,\mathrm{N_{-u}^\downarrow} 
  - \mathrm{Coul_{0t}}\,\mathrm{D_{0d}}\,\overline{\mathrm{Coul_{0d}}}\,\overline{\mathrm{D_{0t}}}\,\mathrm{N_{-u}^\uparrow}
   - \mathrm{C_{0t}}\,\mathrm{Coul_{0d}}\,\overline{\mathrm{C_{0d}}}\,\overline{\mathrm{Coul_{0t}}}\,\overline{\mathrm{N_{-u}^\downarrow}}
   \right.  \nonumber \\
  &\qquad \left.
  -\mathrm{Coul_{0d}}\,\mathrm{D_{0t}}\,\overline{\mathrm{Coul_{0t}}}\,\overline{\mathrm{D_{0d}}}\,\overline{\mathrm{N_{-u}^\uparrow}}\right)\,\mathrm{\tilde{\rho}_{\downarrow \downarrow}}
  \nonumber \\
  +&\left(\mathrm{A_{0t}}\,\mathrm{Coul_{0d}}\overline{\mathrm{A_{0d}}}\,\overline{\mathrm{Coul_{0t}}}\,\mathrm{G_{-u}^\downarrow}\,
  +\mathrm{B_{0t}}\,\mathrm{Coul_{0d}}\,\overline{\mathrm{B_{0d}}}\,\overline{\mathrm{Coul_{0t}}}\,\mathrm{G_{-u}^\uparrow} 
  + \mathrm{C_{0t}}\,\mathrm{Coul_{0d}}\,\overline{\mathrm{C_{0d}}}\,\overline{\mathrm{Coul_{0t}}} \,\mathrm{G_{-u}^\downarrow}
  \right.  \nonumber \\
  &\qquad \left.
   + \mathrm{Coul_{0d}}\,\mathrm{D_{0t}} \,\overline{\mathrm{Coul_{0t}}}\,\overline{\mathrm{D_{0d}}}\,\mathrm{G_{-u}^\uparrow}
   +\mathrm{A_{0d}}\,\mathrm{Coul_{0t}}\,\overline{\mathrm{A_{0t}}}\,\overline{\mathrm{Coul_{0d}}}\,\overline{\mathrm{G_{-u}^\downarrow}}
   +\mathrm{B_{0d}}\,\mathrm{Coul_{0t}}\,\overline{\mathrm{B_{0t}}}\,\overline{\mathrm{Coul_{0d}}}\,\overline{\mathrm{G_{-u}^\uparrow}}
   \right.  \nonumber \\
  &\qquad \left.
  +\mathrm{C_{0d}}\,\mathrm{Coul_{0t}}\,\overline{\mathrm{C_{0t}}}\,\overline{\mathrm{Coul_{0d}}}\,\overline{\mathrm{G_{-u}^\downarrow}}
   +\mathrm{Coul_{0t}}\,\mathrm{D_{0d}}\,\overline{\mathrm{Coul_{0d}}}\,\overline{\mathrm{D_{0t}}}\,\overline{\mathrm{G_{-u}^\uparrow}}\right)\,\mathrm{\tilde{\rho}_{bb}}
\end{align}

The off diagonal element 2,3 yields:
\begin{align}
  -\frac{d\rho_{\uparrow, \downarrow}}{dt} =  \frac{\lambda^2}{\hbar^2} \sum \int du
  & \left(-\mathrm{A_{0t}}\,\mathrm{N_{-u}^\uparrow}\,\overline{\mathrm{C_{0d}}}-\mathrm{B_{0t}}\,\mathrm{N_{-u}^\downarrow}\,\overline{\mathrm{D_{0d}}}-\mathrm{A_{0d}}\,\overline{\mathrm{C_{0t}}}\,\overline{\mathrm{N_{-u}^\uparrow}}-\mathrm{B_{0d}}\,\overline{\mathrm{D_{0t}}}\,\overline{\mathrm{N_{-u}^\downarrow}}\right)\,\tilde{\rho}_{00}
  \nonumber \\
  +&\left(\mathrm{A_{0d}}\,\mathrm{G_{-u}^\uparrow}\,\overline{\mathrm{C_{0t}}}+\mathrm{B_{0d}}\,\mathrm{G_{-u}^\downarrow}\,\overline{\mathrm{D_{0t}}}+\mathrm{A_{0d}}\,\mathrm{Coul_{0t}}\,\mathrm{N_{-u}^\downarrow}\,\overline{\mathrm{C_{0t}}}\,\overline{\mathrm{Coul_{0d}}}+\mathrm{B_{0d}}\,\mathrm{Coul_{0t}}\,\mathrm{N_{-u}^\uparrow}\,\overline{\mathrm{Coul_{0d}}}\,\overline{\mathrm{D_{0t}}}\right)\,\mathrm{\tilde{\rho}_{\uparrow \uparrow}}
  \nonumber \\
  +&\left(\mathrm{C_{0d}}\,\mathrm{G_{-u}^\uparrow}\,\overline{\mathrm{C_{0t}}}+\mathrm{D_{0d}}\,\mathrm{G_{-u}^\downarrow}\,\overline{\mathrm{D_{0t}}}
  +\mathrm{A_{0t}}\,\overline{\mathrm{A_{0d}}}\,\overline{\mathrm{G_{-u}^\uparrow}}
  +\mathrm{B_{0t}}\,\overline{\mathrm{B_{0d}}}\,\overline{\mathrm{G_{-u}^\downarrow}}
  +\mathrm{C_{0d}}\,\mathrm{Coul_{0t}}\,\mathrm{N_{-u}^\downarrow}\,\overline{\mathrm{C_{0t}}}\,\overline{\mathrm{Coul_{0d}}}
  \right.  \nonumber \\
  &\hspace{1.5em} \left.
  +\mathrm{Coul_{0t}}\,\mathrm{D_{0d}}\,\mathrm{N_{-u}^\uparrow}\,\overline{\mathrm{Coul_{0d}}}\,\overline{\mathrm{D_{0t}}}
  +\mathrm{A_{0t}}\,\mathrm{Coul_{0d}}\,\overline{\mathrm{A_{0d}}}\,\overline{\mathrm{Coul_{0t}}}\,\overline{\mathrm{N_{-u}^\downarrow}}
  +\mathrm{B_{0t}}\,\mathrm{Coul_{0d}}\,\overline{\mathrm{B_{0d}}}\,\overline{\mathrm{Coul_{0t}}}\,\overline{\mathrm{N_{-u}^\uparrow}}\right)\,\mathrm{\tilde{\rho}_{\uparrow \downarrow}}
  \nonumber \\
  +&\left(\mathrm{A_{0t}}\,\overline{\mathrm{C_{0d}}}\,\overline{\mathrm{G_{-u}^\uparrow}}+\mathrm{B_{0t}}\,\overline{\mathrm{D_{0d}}}\,\overline{\mathrm{G_{-u}^\downarrow}}+\mathrm{A_{0t}}\,\mathrm{Coul_{0d}}\,\overline{\mathrm{C_{0d}}}\,\overline{\mathrm{Coul_{0t}}}\,\overline{\mathrm{N_{-u}^\downarrow}}+\mathrm{B_{0t}}\,\mathrm{Coul_{0d}}\,\overline{\mathrm{Coul_{0t}}}\,\overline{\mathrm{D_{0d}}}\,\overline{\mathrm{N_{-u}^\uparrow}}\right)\,\mathrm{\tilde{\rho}_{\downarrow \downarrow}}
  \nonumber \\
  +&\left(-\mathrm{A_{0t}}\,\mathrm{Coul_{0d}}\,\mathrm{G_{-u}^\downarrow}\,\overline{\mathrm{C_{0d}}}\,\overline{\mathrm{Coul_{0t}}}
  -\mathrm{B_{0t}}\,\mathrm{Coul_{0d}}\,\mathrm{G_{-u}^\uparrow}\,\overline{\mathrm{Coul_{0t}}}\,\overline{\mathrm{D_{0d}}}-\mathrm{A_{0d}}\,\mathrm{Coul_{0t}}\,\overline{\mathrm{C_{0t}}}\,\overline{\mathrm{Coul_{0d}}}\,\overline{\mathrm{G_{-u}^\downarrow}}
 \right.  \nonumber \\
  &\qquad \left.
   -\mathrm{B_{0d}}\,\mathrm{Coul_{0t}}\,\overline{\mathrm{Coul_{0d}}}\,\overline{\mathrm{D_{0t}}}\,\overline{\mathrm{G_{-u}^\uparrow}}\right)\,\mathrm{\tilde{\rho}_{bb}}
\label{eq:UpDnCouples}
\end{align}
\section{Coupled equations in terms of bath correlation functions}
\label{App:ResultsGN}

Here we state the full coupled differential equations for the density matrix of the system with Fourier transformed reservoir correlation functions, corresponding to Eq. \ref{eq:rhoupupGNGammaNu}.  
 \begin{align}
-\frac{d\tilde{\rho}_{0,0}}{dt} = \frac{\lambda^2}{\hbar^2} \sum_l   & \tilde{\rho}_{00} \left( 
 \left(  \frac{\Omega-a}{2\Omega}  \right) N_{l\uparrow}(\omega_{+-})  + \left( \frac{\Omega+a}{2\Omega}  \right) N_{l\uparrow}(\omega_{++})
   +\left( \frac{\Omega+a}{2\Omega} \right)  N_{l\downarrow}( \omega_{--} )+  \left( \frac{\Omega-a}{2\Omega} \right)   
   N_{l\downarrow}(\omega_{-+}) 
 \right) 
\nonumber \\
&  -\tilde{\rho}_{\uparrow \uparrow} \left(
  \left( \frac{\omega_1}{2\Omega}\right)^2  \left( G_{l\downarrow}( -\omega_{--} ) +  G_{l\downarrow}(-\omega_{-+} )  \right)
+\left(  \frac{\Omega-a}{2\Omega}  \right)^2 G_{l\uparrow}( -\omega_{+-})  + \left( \frac{\Omega+a}{2\Omega} \right)^2   G_{l\uparrow}( -\omega_{++}) 
 \right) 
\nonumber \\
& -\tilde{\rho}_{\downarrow \uparrow} \left( -\frac{\omega_1}{2\Omega} \left( \frac{\Omega+a}{2\Omega}  G_{l\downarrow}(- \omega_{--})- \frac{\Omega-a}{2\Omega}  G_{l\downarrow}(-\omega_{-+}) 
+ \frac{\Omega-a}{2\Omega}  G_{l\uparrow}(-\omega_{+-}) - \frac{\Omega+a}{2\Omega}  G^*_{l\uparrow}(-\omega_{++})\right)
 \right)  \nonumber
\\
&   -\tilde{\rho}_{\uparrow \downarrow} \left(
 -\frac{\omega_1}{2\Omega} \left( \frac{\Omega+a}{2\Omega}  G_{l\downarrow}(-\omega_{--})-  \frac{\Omega-a}{2\Omega}  G_{l\downarrow}(-\omega_{-+})
+  \frac{\Omega-a}{2\Omega}  G_{l\uparrow}(-\omega_{+-}) 
 -   \frac{\Omega+a}{2\Omega}  G_{l\uparrow}(-\omega_{+-}) \right)
 \right) 
\nonumber \\
& -\tilde{\rho}_{\downarrow \downarrow} \left(
  \left( \frac{\Omega+a}{2\Omega} \right)^2    G_{l\downarrow}(- \omega_{--}) +  \left( \frac{\Omega-a}{2\Omega} \right)^2    G_{l\downarrow}(- \omega_{-+})  
 + \left( \frac{\omega_1}{2\Omega}\right)^2 (  G_{l\uparrow}( -\omega_{+-} ) + G_{l\uparrow}( -\omega_{++} ))
 \right) 
 \end{align}

\begin{align}
-\frac{d\tilde{\rho}_{\uparrow \uparrow}}{dt}=   \frac{\lambda^2}{\hbar^2} \sum_l  & \mathrm{\tilde{\rho}_{\uparrow \uparrow}}
\left( 
\left(  \frac{\Omega-a}{2\Omega}  \right)^2 G_{l\uparrow}(-\omega_{+-})  + \left( \frac{\Omega+a}{2\Omega}  \right)^2   G_{l\uparrow}(-\omega_{++})
+\left( \frac{\omega_1}{2\Omega}\right)^2  \left(N_{l\uparrow}(\frac{E_B}{\hbar}-\omega_{--}) + N_{l\uparrow}(\frac{E_B}{\hbar}-\omega_{-+} )  \right)
\right. \nonumber \\
&\qquad \left.  
+ \left( \frac{\omega_1}{2\Omega}\right)^2  \left( G_{l\downarrow}(-\omega_{--}) + G_{l\downarrow}(-\omega_{-+})  \right)
+ \left(  \frac{\Omega-a}{2\Omega}  \right)^2 N_{l\downarrow}(\frac{E_B}{\hbar}-\omega_{+-})  + \left( \frac{\Omega+a}{2\Omega}  \right)^2   N_{l\downarrow}(\frac{E_B}{\hbar}-\omega_{++}) 
\right) \nonumber \\
+& \mathrm{\tilde{\rho}_{\uparrow \downarrow}}
\left( -\frac{\omega_1}{2\Omega} \right)
 \left( \frac{\Omega-a}{2\Omega} 
\left( \overline{\Gamma_{l\uparrow}(-\omega_{+-})} +\overline{\nu_{l\downarrow}(\frac{E_B}{\hbar}-\omega_{+-}) }
- \overline{\nu_{l\uparrow}(\frac{E_B}{\hbar}-\omega_{-+})} -\overline{\Gamma_{l\downarrow}(-\omega_{-+})} \right)
\right. \nonumber \\
& \qquad \qquad \qquad \left. 
- \frac{\Omega+a}{2\Omega} 
\left( \overline{\Gamma_{l\uparrow}(-\omega_{++})} +\overline{\nu_{l\downarrow}(\frac{E_B}{\hbar}-\omega_{++})}
-
  \overline{\nu_{l\uparrow}(\frac{E_B}{\hbar}-\omega_{--})}  -\overline{\Gamma_{l\downarrow}(-\omega_{--})} \right) 
 \right) 
 \nonumber \\
+& \mathrm{\tilde{\rho}_{\downarrow \uparrow}}
\left(
\left( -\frac{\omega_1}{2\Omega} \right) \left( \frac{\Omega-a}{2\Omega} 
\left( \Gamma_{l\uparrow}(-\omega_{+-}) +\nu_{l\downarrow}(\frac{E_B}{\hbar}-\omega_{+-}) 
-\nu_{l\uparrow}(\frac{E_B}{\hbar}-\omega_{-+}) -\Gamma_{l\downarrow}(-\omega_{-+})
 \right)
 \right. \right. \nonumber \\
&\qquad \qquad \qquad \left. \left.
  + 
 \frac{\Omega+a}{2\Omega}  \left( \nu_{l\uparrow}(\frac{E_B}{\hbar}-\omega_{--}) +\Gamma_{l\downarrow}(-\omega_{--}) 
- \Gamma_{l\uparrow}(-\omega_{++}) -\nu_{l\downarrow}(\frac{E_B}{\hbar}-\omega_{++})
\right)    \right)
\right)
 \nonumber \\
-& \mathrm{\tilde{\rho}_{00}}
\left(
 \left(  \frac{\Omega-a}{2\Omega}  \right)^2 N_{l\uparrow}(\omega_{+-})  + \left( \frac{\Omega+a}{2\Omega}  \right)^2   N_{l\uparrow}(\omega_{++})  
+\left( \frac{\omega_1}{2\Omega}\right)^2   \left( N_{l\downarrow}(\omega_{--}) + N_{l\downarrow}(\omega_{-+})  \right)
\right)
 \nonumber \\
-& \mathrm{\tilde{\rho}_{bb}}
\left(
 \left( \frac{\omega_1}{2\Omega}\right)^2  \left( G_{l\uparrow}(\omega_{--}-\frac{E_B}{\hbar}) + G_{l\uparrow}(\omega_{-+}-\frac{E_B}{\hbar})  \right) 
\right. 
\nonumber \\
& \qquad \left.
+  \left(  \frac{\Omega-a}{2\Omega}  \right)^2 G_{l\downarrow}(\omega_{+-}-\frac{E_B}{\hbar})  + \left( \frac{\Omega+a}{2\Omega}  \right)^2   G_{l\downarrow}(\omega_{++}-\frac{E_B}{\hbar})  
\right)
\end{align}

\begin{align}
-\frac{d\tilde{\rho}_{\downarrow, \downarrow }}{dt} =   \frac{\lambda^2}{\hbar^2} \sum_l &\tilde{\rho}_{\downarrow \downarrow}\left (
\left( \frac{\omega_1}{2\Omega}\right)^2  (G_{l,\uparrow}(-\omega_{+-}) +G_{l,\uparrow}(-\omega_{++} )) 
+ \left( \frac{\Omega+a}{2\Omega} \right)^2   G_{l,\downarrow}(- \omega_{--}) +  \left( \frac{\Omega-a}{2\Omega} \right)^2    G_{l,\downarrow}(-\omega_{-+} ) 
\right. \nonumber \\
&+ \left.  \left( \frac{\Omega+a}{2\Omega} \right)^2   N_{l\uparrow}(\frac{E_B}{\hbar} - \omega_{--}) +  \left( \frac{\Omega-a}{2\Omega} \right)^2   N_{l\uparrow}(\frac{E_B}{\hbar}- \omega_{-+} ) 
\right. 
\nonumber \\
 &+ \left.
\left( \frac{\omega_1}{2\Omega}\right)^2  \left( N_{l\downarrow}(\frac{E_B}{\hbar}-\omega_{+-}) + N_{l\downarrow}(\frac{E_B}{\hbar}-\omega_{++} )\right)
\right) 
\nonumber \\
	+&\tilde{\rho}_{\uparrow \downarrow} \left( -\frac{\omega_1}{2\Omega} \right) \left( \frac{\Omega-a}{2\Omega}   \left(\Gamma_{l,\uparrow}(-\omega_{+-})+\nu_{l\downarrow}(\frac{E_B}{\hbar}-\omega_{+-}) 
	-\Gamma_{l,\downarrow}(-\omega_{-+}) - \nu_{l\uparrow}(\frac{E_B}{\hbar}-\omega_{-+}) \right)  
\right. \nonumber \\
&\qquad \qquad \left.
	+ \frac{\Omega+a}{2\Omega} (\Gamma_{l,\downarrow}(-\omega_{--})+\nu_{l\uparrow}(\frac{E_B}{\hbar}-\omega_{--})  
	- \Gamma_{l,\uparrow}(-\omega_{++}) - \nu_{l\downarrow}(\frac{E_B}{\hbar}-\omega_{++}) ) 
\right)
\nonumber \\
	+&\tilde{\rho}_{\downarrow \uparrow} \left(  -\frac{\omega_1}{2\Omega} \right) \left( \frac{\Omega-a}{2\Omega}  (\overline{\Gamma_{l,\uparrow}(-\omega_{+-})}+ \overline{\nu_{l\downarrow}(\frac{E_B}{\hbar}-\omega_{+-})}
	 - \overline{\Gamma_{l,\downarrow}(-\omega_{-+}) }-  \overline{\nu_{l\uparrow}(\frac{E_B}{\hbar}-\omega_{-+})} )
\right. \nonumber \\
& \qquad \qquad \left.
	 + \frac{\Omega+a}{2\Omega} 
	\left( \overline{\Gamma_{l,\downarrow}(-\omega_{--})} + \overline{\nu_{l\uparrow}(\frac{E_B}{\hbar}-\omega_{--})}
	-\overline{\Gamma_{l,\uparrow}(-\omega_{++})} - \overline{\nu_{l\downarrow}(\frac{E_B}{\hbar}-\omega_{++})}  \right) 
\right)
\nonumber \\
	-&\tilde{\rho}_{00} \left(  \left( \frac{\Omega+a}{2\Omega} \right)^2   N_{l\downarrow}(\omega_{--}) +  \left( \frac{\Omega-a}{2\Omega} \right)^2    N_{l\downarrow}(\omega_{-+})  
+\left( \frac{\omega_1}{2\Omega}\right)^2  ( N_{l\uparrow}( \omega_{+-}) +  N_{l\uparrow}( \omega_{++} ))
\right)
\nonumber \\
	-&\tilde{\rho}_{bb} \left(  \left( \frac{\Omega+a}{2\Omega} \right)^2   G_{l,\uparrow}( \omega_{--}-\frac{E_B}{\hbar} )+  \left( \frac{\Omega-a}{2\Omega} \right)^2   G_{l,\uparrow}(\omega_{-+}-\frac{E_B}{\hbar}) 
\right.
\nonumber \\
& \qquad+
\left.
 \left( \frac{\omega_1}{2\Omega}\right)^2  (G_{l,\downarrow}(\omega_{+-}-\frac{E_B}{\hbar}) + G_{l,\downarrow}(\omega_{++}-\frac{E_B}{\hbar} ))
\right)
\end{align}

\begin{align}
-\frac{d \tilde{\rho}_{\uparrow \downarrow}}{dt} =  \frac{\lambda^2}{\hbar^2}  \sum_l 
&\tilde{\rho}_{\uparrow \downarrow} \left(
 \left( \frac{\omega_1}{2\Omega}\right)^2  ( \overline{\Gamma_{l\uparrow}( -\omega_{+-})} +  \overline{\Gamma_{l\uparrow}(-\omega_{++})} )
+ \left(  \frac{\Omega-a}{2\Omega}  \right)^2 \Gamma_{l\uparrow}(- \omega_{+-})  + \left( \frac{\Omega+a}{2\Omega}  \right)^2   \Gamma_{l\uparrow}(-\omega_{++}) 
\right. \nonumber \\
& \left. \qquad
+  \left( \frac{\Omega+a}{2\Omega} \right)^2    \overline{\Gamma_{l\downarrow}( - \omega_{--})} +  \left( \frac{\Omega-a}{2\Omega} \right)^2   \overline{\Gamma_{l\downarrow}( -\omega_{-+} )} 
+ \left( \frac{\omega_1}{2\Omega}\right)^2  \left( \Gamma_{l\downarrow}(-\omega_{--}) + \Gamma_{l\downarrow}(-\omega_{-+})  \right)
\right. \nonumber \\
& \left. \qquad+  \left( \frac{\Omega+a}{2\Omega} \right)^2   \overline{\nu_{l\uparrow}(\frac{E_B}{\hbar} - \omega_{--})}  +  \left( \frac{\Omega-a}{2\Omega} \right)^2  \overline{\nu_{l\uparrow}(\frac{E_B}{\hbar}-  \omega_{-+}) } 
\right. \nonumber \\
& \left. \qquad
 + \left( \frac{\omega_1}{2\Omega}\right)^2  \left( \nu_{l\uparrow}(\frac{E_B}{\hbar} -\omega_{--})  + \nu_{l\uparrow}(\frac{E_B}{\hbar} -\omega_{-+} )  \right)
\right. \nonumber \\
& \left. \qquad
+ \left( \left( \frac{\omega_1}{2\Omega}\right)^2  ( \overline{\nu_{l\downarrow}(\frac{E_B}{\hbar} -\omega_{+-})} +\overline{\nu_{l\downarrow}(\frac{E_B}{\hbar} -\omega_{++})} \right)
\right. \nonumber \\
& \left. \qquad
+ \left( \left(  \frac{\Omega-a}{2\Omega}  \right)^2 \nu_{l\downarrow}(\frac{E_B}{\hbar} -\omega_{+-})  + \left( \frac{\Omega+a}{2\Omega}  \right)^2   \nu_{l\downarrow}(\frac{E_B}{\hbar}- \omega_{++})  \right)
\right)
\nonumber \\
+&\tilde{\rho}_{\uparrow \uparrow}\left(
\frac{-\omega_1}{2h}\right) \left( \frac{\Omega-a}{2\Omega}  \left( \overline{\Gamma_{l\uparrow}(-\omega_{+-})} + \overline{\nu_{l\downarrow}(\frac{E_B}{\hbar}-\omega_{+-} )}
- \overline{\Gamma_{l\downarrow}(-\omega_{-+})}- \overline{\nu_{l\uparrow}(\frac{E_B}{\hbar}-\omega_{-+} )}  \right)  
\right. \nonumber \\
& \left. \qquad \qquad
+ \frac{\Omega+a}{2\Omega}  \left( \overline{\Gamma_{l\downarrow}(-\omega_{--})}  + \overline{\nu_{l\uparrow}(\frac{E_B}{\hbar}-\omega_{--} )} -  \overline{\Gamma_{l\uparrow}(-\omega_{++})}  - \overline{\nu_{l\downarrow}(\frac{E_B}{\hbar}-\omega_{++} )}  \right)    
\right)
\\
+&\tilde{\rho}_{\downarrow \downarrow} \left(
 \left(-\frac{\omega_1}{2\Omega}\right)\left( \frac{\Omega-a}{2\Omega}    (\Gamma_{l\uparrow}(-\omega_{+-})+ \nu_{l\downarrow}( \frac{E_B}{\hbar}-\omega_{+-}) - \Gamma_{l\downarrow}(-\omega_{-+})- \nu_{l\uparrow}(\frac{E_B}{\hbar}-\omega_{-+} ) ) 
\right. \right.  \nonumber \\
& \left.  \left. \qquad \qquad
+  \frac{\Omega+a}{2\Omega}  \left( \Gamma_{l\downarrow}(-\omega_{--})+ \nu_{l\uparrow}(\frac{E_B}{\hbar} -\omega_{--})  - \Gamma_{l\uparrow}(-\omega_{++}) - \nu_{l\downarrow}(\frac{E_B}{\hbar}- \omega_{++}) \right) \right) 
 \right)
\nonumber \\
-&\tilde{\rho}_{00} \left(
 \left(-\frac{\omega_1}{2\Omega}\right)\left( \frac{\Omega-a}{2\Omega}  N_{l\uparrow}( \omega_{+-})  
- \frac{\Omega+a}{2\Omega}  N_{l\uparrow}(\omega_{++})
 + \frac{\Omega+a}{2\Omega}  N_{l\downarrow}(  \omega_{--})
 - \frac{\Omega-a}{2\Omega}  N_{l\downarrow}( \omega_{-+}) 
  \right) 
\right)
\nonumber \\
-&\tilde{\rho}_{bb} \left(
  -\frac{\omega_1}{2\Omega}\right) \left( \frac{\Omega+a}{2\Omega}  \left( G_{l\uparrow}(\omega_{--}-\frac{E_B}{\hbar}) - G_{l\downarrow}(\omega_{++}-\frac{E_B}{\hbar}) \right)
 -  \frac{\Omega-a}{2\Omega} \left( G_{l\uparrow}(\omega_{-+}-\frac{E_B}{\hbar}) 
-  G_{l\downarrow}(\omega_{+-}-\frac{E_B}{\hbar}) \right)
  \right) 
\nonumber
\end{align}

\begin{align}
-\frac{d\tilde{\rho}_{b,b}}{dt} =   \frac{\lambda^2}{\hbar^2} \sum_l  \, \tilde{\rho}_{bb} & \left(
\left( \frac{\omega_1}{2\Omega}\right)^2  \left(G_{l\uparrow}(\omega_{--}-\frac{E_B}{\hbar})+ G_{l\uparrow}(\omega_{-+}-\frac{E_B}{\hbar})  \right)
+\left(  \frac{\Omega+a}{2\Omega} \right)^2    G_{l\uparrow}( \omega_{--}-\frac{E_B}{\hbar})
\right. \nonumber \\
& \, \left.
 +  \left( \frac{\Omega-a}{2\Omega} \right)^2   G_{l\uparrow}(\omega_{-+} -\frac{E_B}{\hbar}) 
+ \left(  \frac{\Omega-a}{2\Omega}  \right)^2  G_{l\downarrow}( \omega_{+-}-\frac{E_B}{\hbar})  + \left( \frac{\Omega+a}{2\Omega}  \right)^2   G_{l\downarrow}(\omega_{++}-\frac{E_B}{\hbar})
\right. \nonumber \\
& \, \left.
+\left( \frac{\omega_1}{2\Omega}\right)^2 
 (G_{l\downarrow}(\omega_{+-}-\frac{E_B}{\hbar}) + G_{l\downarrow}(\omega_{++}-\frac{E_B}{\hbar}))
\right) \nonumber \\
-\tilde{\rho}_{\uparrow \uparrow} &\left(
\left( \frac{\omega_1}{2\Omega}\right)^2  \left( N_{l\uparrow}(\frac{E_B}{\hbar}-\omega_{--}) + N_{l\uparrow}(\frac{E_B}{\hbar}-\omega_{-+} )  \right)
\right. \nonumber \\
& \, \left.
+ \left(  \frac{\Omega-a}{2\Omega}  \right)^2  N_{l\downarrow}(\frac{E_B}{\hbar}-\omega_{+-})  + \left( \frac{\Omega+a}{2\Omega}  \right)^2   N_{l\downarrow}(\frac{E_B}{\hbar}-\omega_{++})\right)
\nonumber \\
-\tilde{\rho}_{\downarrow \downarrow} &\left( \left( \frac{\Omega+a}{2\Omega} \right)^2 
N_{l\uparrow}(\frac{E_B}{\hbar}- \omega_{--}) +  \left( \frac{\Omega-a}{2\Omega} \right)^2  N_{l\uparrow}(\frac{E_B}{\hbar} -\omega_{-+}) 
\right. \nonumber \\
& \, \left.
+\left( \frac{\omega_1}{2\Omega}\right)^2 
 (N_{l\downarrow}(\frac{E_B}{\hbar}-\omega_{+-} ) + N_{l\downarrow}(\frac{E_B}{\hbar}-\omega_{++})) \right)
\nonumber \\
+\tilde{\rho}_{\uparrow \downarrow} &\left(
 -\frac{\omega_1}{2\Omega} \left( \frac{\Omega+a}{2\Omega} N_{l\uparrow}(\frac{E_B}{\hbar}-\omega_{--}) - \frac{\Omega-a}{2\Omega} N_{l\uparrow}(\frac{E_B}{\hbar}-\omega_{-+}) \right)
 \right.  \nonumber \\
  &\qquad \left.
-\frac{\omega_1}{2\Omega} \left( \frac{\Omega-a}{2\Omega}  N_{l\downarrow}(\frac{E_B}{\hbar}-\omega_{+-}) - \frac{\Omega+a}{2\Omega}  N_{l\downarrow}(\frac{E_B}{\hbar}-\omega_{++})\right)
 \right)
\nonumber \\
+\tilde{\rho}_{\downarrow \uparrow} &\left(
 -\frac{\omega_1}{2\Omega} \left( \frac{\Omega+a}{2\Omega}  N_{l\uparrow}(\frac{E_B}{\hbar}-\omega_{--})  - \frac{\Omega-a}{2\Omega}  N_{l\uparrow}(\frac{E_B}{\hbar}-\omega_{-+})   \right)
 \right.  \nonumber \\
  &\qquad \left.
-\frac{\omega_1}{2\Omega} \left( \frac{\Omega-a}{2\Omega}  N_{l\downarrow}(\frac{E_B}{\hbar}-\omega_{+-}) - \frac{\Omega+a}{2\Omega}  N_{l\uparrow}(\frac{E_B}{\hbar}-\omega_{++}) \right) \right)
\end{align}

We omit for brevity the other coherences which do not couple with the terms above. 

\section{Results}
\label{App:Results}
Below are the full coupled differential equations for the elements of the density matrix that couple with the populations (other coherences simply decay away, and are generally initially zero anyway).
\begin{align}
-\frac{d\tilde{\rho}_{0,0}}{dt} &=  \frac{2\pi \lambda^2 D}{\hbar^2} \tilde{\rho}_{00} 
\left( 
 \frac{\Omega-a}{2\Omega} \left(n_e(\hbar \omega_{+-})+ n_e(\hbar \omega_{-+})\right)
   + \frac{\Omega+a}{2\Omega}  \left( n_e(\hbar \omega_{++}) + n_e(\hbar \omega_{--})\right)
 \right) 
\nonumber \\
 & -\frac{2\pi \lambda^2 D}{\hbar^2}\tilde{\rho}_{\uparrow \uparrow}
\left(
\left(\frac{\omega_1}{2\Omega}\right)^2 \left( n_h(\hbar \omega_{--})+ n_h(\hbar \omega_{-+})\right)
+ \left( \frac{\Omega-a}{2\Omega} \right)^2 n_h(\hbar \omega_{+-})
+ \left( \frac{\Omega+a}{2\Omega} \right)^2 n_h(\hbar \omega_{++})
\right)  
\nonumber \\
& -\frac{2\pi \lambda^2 D}{\hbar^2} \tilde{\rho}_{\downarrow \uparrow} \left( 
\left(-\frac{\omega_1}{2\Omega}\right) \left( \frac{\Omega+a}{2\Omega} \right)\left(n_h( \hbar \omega_{--})- n_h(\hbar \omega_{++}) \right)
+\left(-\frac{\omega_1}{2\Omega}\right) \left( \frac{\Omega-a}{2\Omega} \right)\left(n_h( \hbar \omega_{+-})- n_h(\hbar \omega_{-+}) \right)
 \right)  \nonumber
\\
&   -\frac{2\pi \lambda^2 D}{\hbar^2} \tilde{\rho}_{\uparrow \downarrow} \left(
 \left(-\frac{\omega_1}{2\Omega}\right) \left( \frac{\Omega+a}{2\Omega} \right)\left(n_h(\hbar \omega_{--}) -n_h(\hbar \omega_{+-}) \right)
+ \left(-\frac{\omega_1}{2\Omega}\right) \left( \frac{\Omega-a}{2\Omega} \right)\left(n_h(\hbar \omega_{+-}) -n_h(\hbar \omega_{-+})\right) 
\right)
\nonumber \\
& -\frac{2\pi \lambda^2 D}{\hbar^2}\tilde{\rho}_{\downarrow \downarrow} \left(
    \left( \frac{\Omega+a}{2\Omega} \right)^2    n_h( \hbar \omega_{--}) +  \left( \frac{\Omega-a}{2\Omega} \right)^2    n_h( \hbar \omega_{-+})  
 + \left( \frac{\omega_1}{2\Omega}\right)^2 (  n_h( \hbar \omega_{+-} ) + n_h(\hbar \omega_{++} ))
\right)
\nonumber \\
 \end{align}

\begin{align}
-\frac{d\tilde{\rho}_{\uparrow \uparrow}}{dt}= \frac{2\pi \lambda^2 D}{\hbar^2}  & \mathrm{\tilde{\rho}_{\uparrow \uparrow}}
\left( 
\left(  \frac{\Omega-a}{2\Omega}  \right)^2 n_h(\hbar \omega_{+-})  + \left( \frac{\Omega+a}{2\Omega}  \right)^2   n_h(\hbar \omega_{++})
+\left( \frac{\omega_1}{2\Omega}\right)^2  \left(n_e(E_b- \hbar \omega_{--}) + n_e(E_b- \hbar \omega_{-+} )  \right)
\right. \nonumber \\
&\qquad \left.  
+ \left( \frac{\omega_1}{2\Omega}\right)^2  \left( n_h(\hbar \omega_{--}) + n_h(\hbar \omega_{-+})  \right)
+ \left(  \frac{\Omega-a}{2\Omega}  \right)^2 n_e(E_b- \hbar \omega_{+-})  + \left( \frac{\Omega+a}{2\Omega}  \right)^2   n_e(E_b- \hbar \omega_{++}) 
\right) \nonumber \\
+& \mathrm{\tilde{\rho}_{\uparrow \downarrow}}
\left( -\frac{\omega_1}{2\Omega} \right)
 \left( \frac{\Omega-a}{2\Omega} 
\left( \frac{1}{2}n_h(\hbar \omega_{+-}) +\frac{1}{2}n_e(E_b- \hbar \omega_{+-}) 
- \frac{1}{2}n_e(E_b- \hbar \omega_{-+}) -\frac{1}{2}n_h(\hbar \omega_{-+}) \right)
\right. \nonumber \\
&\qquad  \qquad\left.
- \frac{\Omega+a}{2\Omega} 
\left( \frac{1}{2}n_h(\hbar \omega_{++}) +\frac{1}{2}n_e(E_b- \hbar \omega_{++})
-
  \frac{1}{2}n_e(E_b- \hbar \omega_{--}) -\frac{1}{2}n_h(\hbar \omega_{--}) \right) 
 \right) 
 \nonumber \\
+& \mathrm{\tilde{\rho}_{\downarrow \uparrow}}
\left(
\left( -\frac{\omega_1}{2\Omega} \right) \left( \frac{\Omega-a}{2\Omega} 
\left( \frac{1}{2}n_h(\hbar \omega_{+-}) +\frac{1}{2}n_e(E_b- \hbar \omega_{+-}) 
-\frac{1}{2}n_e(E_b- \hbar \omega_{-+}) -\frac{1}{2}n_h(\hbar \omega_{-+})
 \right)
\right. \right. \nonumber \\
&\qquad  \qquad \left. \left.
  + 
 \frac{\Omega+a}{2\Omega}  \left( \frac{1}{2}n_e(E_b- \hbar \omega_{--}) +\frac{1}{2}n_h(\hbar \omega_{--}) 
- \frac{1}{2}n_h(\hbar \omega_{++}) -\frac{1}{2}n_e(E_b- \hbar \omega_{++})
\right)    \right)
\right)
 \nonumber \\
-& \mathrm{\tilde{\rho}_{00}}
\left(
 \left(  \frac{\Omega-a}{2\Omega}  \right)^2 n_e(\hbar \omega_{+-})  + \left( \frac{\Omega+a}{2\Omega}  \right)^2   n_e(\hbar \omega_{++})  
+\left( \frac{\omega_1}{2\Omega}\right)^2   \left( n_e(\hbar \omega_{--}) +n_e(\hbar \omega_{-+})  \right)
\right)
 \nonumber \\
-& \mathrm{\tilde{\rho}_{bb}}
\left(
 \left( \frac{\omega_1}{2\Omega}\right)^2  \left( n_h(-\hbar \omega_{--}+E_b) + n_h(-\hbar \omega_{-+}+E_b)  \right) 
\right.  \nonumber \\
  &\qquad \left.
+  \left( \left(  \frac{\Omega-a}{2\Omega}  \right)^2 n_h(-\hbar \omega_{+-}+E_b)  + \left( \frac{\Omega+a}{2\Omega}  \right)^2   n_h(-\hbar \omega_{++}+E_b)  \right)
\right)
\end{align}

\begin{align}
-\frac{d\tilde{\rho}_{\downarrow, \downarrow }}{dt} = \frac{2\pi \lambda^2 D}{\hbar^2}
&\tilde{\rho}_{\downarrow \downarrow}\left (
\left( \frac{\omega_1}{2\Omega}\right)^2  (n_h(\hbar \omega_{+-}) +n_h(\hbar \omega_{++} )) 
+ \left( \frac{\Omega+a}{2\Omega} \right)^2  n_h( \hbar \omega_{--}) +  \left( \frac{\Omega-a}{2\Omega} \right)^2   n_h(\hbar \omega_{-+} ) 
\right. \nonumber \\
&+ \left.  \left( \frac{\Omega+a}{2\Omega} \right)^2   n_e(E_b - \hbar \omega_{--}) +  \left( \frac{\Omega-a}{2\Omega} \right)^2   n_e(E_b- \hbar \omega_{-+} ) 
+
\left( \frac{\omega_1}{2\Omega}\right)^2  \left( n_e(E_b- \hbar \omega_{+-}) + n_e(E_b- \hbar \omega_{++} )\right)
\right) 
\nonumber \\
	+&\tilde{\rho}_{\uparrow \downarrow} \left( -\frac{\omega_1}{2\Omega} \right) \left( \frac{\Omega-a}{2\Omega}   \left(\frac{1}{2}n_h(\hbar \omega_{+-})+\frac{1}{2}n_e(E_b- \hbar \omega_{+-}) 
	-\frac{1}{2}n_h(\hbar \omega_{-+}) - \frac{1}{2}n_e(E_b- \hbar \omega_{-+}) \right)  
\right. \nonumber \\
&\qquad \qquad \left.
	+ \frac{\Omega+a}{2\Omega} (\frac{1}{2}n_h(\hbar \omega_{--})+\frac{1}{2}n_e(E_b- \hbar \omega_{--})  
	- \frac{1}{2}n_h(\hbar \omega_{++}) - \frac{1}{2}n_e(E_b- \hbar \omega_{++}) ) 
\right)
\nonumber \\
	+&\tilde{\rho}_{\downarrow \uparrow} \left(  -\frac{\omega_1}{2\Omega} \right) \left( \frac{\Omega-a}{2\Omega}  (\frac{1}{2}n_h(\hbar \omega_{+-})+ \frac{1}{2}n_e(E_b- \hbar \omega_{+-})
	 - \frac{1}{2}n_h(\hbar \omega_{-+}) - \frac{1}{2}n_e(E_b- \hbar \omega_{-+}) )
\right. \nonumber \\
& \qquad \qquad \left.
	 + \frac{\Omega+a}{2\Omega} 
	\left(\frac{1}{2}n_h(\hbar \omega_{--}) + \frac{1}{2}n_e(E_b- \hbar \omega_{--})
	-\frac{1}{2}n_h(\hbar \omega_{++}) - \frac{1}{2}n_e(E_b- \hbar \omega_{++})  \right) 
\right)
\nonumber \\
	-&\tilde{\rho}_{00} \left(  \left( \frac{\Omega+a}{2\Omega} \right)^2   n_e(\hbar \omega_{--}) +  \left( \frac{\Omega-a}{2\Omega} \right)^2   n_e(\hbar \omega_{-+})  
+\left( \frac{\omega_1}{2\Omega}\right)^2  ( n_e( \hbar \omega_{+-}) + n_e( \hbar \omega_{++} ))
\right)
\nonumber \\
	-&\tilde{\rho}_{bb} \left(  \left( \frac{\Omega+a}{2\Omega} \right)^2  n_h(-\hbar \omega_{--}+E_b )+  \left( \frac{\Omega-a}{2\Omega} \right)^2   n_h(-\hbar \omega_{-+}+E_b) 	
\right. \nonumber \\
& \, \left. \qquad
	+ \left( \frac{\omega_1}{2\Omega}\right)^2  (n_h(-\hbar \omega_{+-}+E_b) + n_h(-\hbar \omega_{++} + E_b ))
\right)
\end{align}

\begin{align}
-\frac{d\tilde{\rho}_{b,b}}{dt} =\frac{2\pi \lambda^2 D}{\hbar^2} \,\biggl( \tilde{\rho}_{bb}\biggr. &\left. \left(
\left( \frac{\omega_1}{2\Omega}\right)^2  \left(n_h(-\hbar \omega_{--}+E_b)+ n_h(-\hbar \omega_{-+}+E_b)  \right)
+\left(  \frac{\Omega+a}{2\Omega} \right)^2    n_h( -\hbar \omega_{--}+E_b) 
\right. \right. \nonumber \\
& \, \left. \left.
+  \left( \frac{\Omega-a}{2\Omega} \right)^2   n_h(-\hbar \omega_{-+} +E_b) 
+ \left(  \frac{\Omega-a}{2\Omega}  \right)^2  n_h( -\hbar \omega_{+-}+E_b)  + \left( \frac{\Omega+a}{2\Omega}  \right)^2   n_h(-\hbar \omega_{++}+E_b)
\right. \right. \nonumber \\
& \, \left. \left.
+\left( \frac{\omega_1}{2\Omega}\right)^2 
n_h(-\hbar \omega_{+-}+E_b) + n_h(-\hbar \omega_{++}+E_b))
\right) \right. \nonumber \\
-\tilde{\rho}_{\uparrow \uparrow} &\left. \left(
\left( \frac{\omega_1}{2\Omega}\right)^2  \left( n_e(E_b- \hbar \omega_{--}) + n_e(E_b- \hbar \omega_{-+} )  \right)
+ \left(  \frac{\Omega-a}{2\Omega}  \right)^2  n_e(E_b- \hbar \omega_{+-})  
\right. \right. \nonumber \\
& \, \left. \left.
+ \left( \frac{\Omega+a}{2\Omega}  \right)^2   n_e(E_b- \hbar \omega_{++})\right)
\right. \nonumber \\
-\tilde{\rho}_{\downarrow \downarrow} & \left. \left( \left( \frac{\Omega+a}{2\Omega} \right)^2 
n_e(E_b-\hbar \omega_{--}) +  \left( \frac{\Omega-a}{2\Omega} \right)^2 n_e(E_b -\hbar\omega_{-+}) 
\right. \right. \nonumber \\
& \, \left. \left.
+\left( \frac{\omega_1}{2\Omega}\right)^2 
 (n_e(E_b- \hbar \omega_{+-} ) + n_e(E_b- \hbar \omega_{++})) \right)
\right. \nonumber \\
+\tilde{\rho}_{\uparrow \downarrow} &\left. \left(
 -\frac{\omega_1}{2\Omega} \left( \frac{\Omega+a}{2\Omega} \left( n_e(E_b- \hbar \omega_{--})- n_e(E_b- \hbar \omega_{++}) \right)
 - \frac{\Omega-a}{2\Omega} \left( n_e(E_b- \hbar \omega_{-+})
- n_e(E_b- \hbar \omega_{+-}) \right)
 \right)
 \right)
\right. \nonumber \\
+\tilde{\rho}_{\downarrow \uparrow} &\left. \left(
 -\frac{\omega_1}{2\Omega} \left( \frac{\Omega+a}{2\Omega}  \left( n_e(E_b- \hbar \omega_{--})  - n_e(E_b- \hbar \omega_{++})\right)  - \frac{\Omega-a}{2\Omega} \left(  n_e(E_b- \hbar \omega_{-+})  - n_e(E_b- \hbar \omega_{+-}) \right) \right) \right)
\right)
\end{align}

\begin{align}
-\frac{d \tilde{\rho}_{\uparrow \downarrow}}{dt} = \frac{2\pi \lambda^2 D}{\hbar^2}
&\tilde{\rho}_{\uparrow \downarrow}\left( 
\frac{h-a}{2} (n_h( \hbar \omega_{+-}) + n_h( \hbar \omega_{-+} )  +   n_e(E_b- \hbar \omega_{-+}) + n_e(E_b - \hbar \omega_{+-}) ) 
\right. \nonumber \\
& \left. \qquad
+ \frac{ h+a}{2}\left( n_h(\hbar \omega_{++}) +n_h( \hbar \omega_{--})  \left( {h+a}\right)  n_e(E_b - \hbar \omega_{--}) 
  + \left( {h+a} \right) n_e(E_b- \hbar \omega_{++}) \right) 
\right)
\nonumber \\
+&\tilde{\rho}_{\uparrow \uparrow}\left(
\frac{-\omega_1}{2h}\right) \left( \frac{\Omega-a}{2\Omega}   \left(\frac{1}{2}n_h(\hbar \omega_{+-})+ \frac{1}{2}n_e(E_b- \hbar \omega_{+-} )
- \frac{1}{2}n_h(\hbar \omega_{-+})- \frac{1}{2}n_e(E_b- \hbar \omega_{-+} ) \right)  
\right. \nonumber \\
& \left. \qquad \qquad
+ \frac{\Omega+a}{2\Omega}  \left(\frac{1}{2}n_h(\hbar \omega_{--})  +\frac{1}{2}n_e(E_b- \hbar \omega_{--} ) -  \frac{1}{2}n_h(\hbar \omega_{++})  - \frac{1}{2}n_e(E_b- \hbar \omega_{++} ) \right)    
\right)
\\
+&\tilde{\rho}_{\downarrow \downarrow} \left(
 -\frac{\omega_1}{2\Omega}\right)\left( \frac{\Omega-a}{2\Omega}    \left(\frac{1}{2}n_h(\hbar \omega_{+-})+ \frac{1}{2}n_e( E_b-\hbar \omega_{+-}) - \frac{1}{2}n_h(\hbar \omega_{-+})-\frac{1}{2}n_e(E_b- \hbar \omega_{-+} ) \right) 
 \right. \nonumber \\
& \left. \qquad \qquad
+  \frac{\Omega+a}{2\Omega}  \left(\frac{1}{2}n_h(\hbar \omega_{--})+\frac{1}{2}n_e(E_b -\hbar \omega_{--})  - \frac{1}{2}n_h(\hbar \omega_{++}) - \frac{1}{2}n_e(E_b- \hbar \omega_{++}) \right)  
 \right)
\nonumber \\
-&\tilde{\rho}_{00} \left(
 \left(-\frac{\omega_1}{2\Omega}\right)\left( \frac{\Omega-a}{2\Omega}  \left( n_e( \hbar \omega_{+-})  - n_e( \hbar \omega_{-+}) \right)
+ \frac{\Omega+a}{2\Omega}  \left(- n_e(\hbar \omega_{++}) + n_e(\hbar  \omega_{--}) \right) \right) 
\right)
\nonumber \\
-&\tilde{\rho}_{bb} \left(
 \left( -\frac{\omega_1}{2\Omega}\right) \left( \frac{\Omega+a}{2\Omega} \left( n_h(-\hbar \omega_{--}+E_b) - n_h(-\hbar \omega_{++}+E_b) \right)
\right. \right.
\nonumber \\
& \qquad \left. \left.
+ \frac{\Omega-a}{2\Omega} \left( n_h(-\hbar \omega_{+-}+E_b)  -  n_h(-\hbar \omega_{-+}+E_b) \right)
\right) 
\right)
\nonumber
\end{align} 

\end{widetext}
 
 \bibliography{MyLibrary}

\begin{thebibliography}{10}

\bibitem{Townsend2025}
Emily Townsend, Joshua Pomeroy, and Garnett~W. Bryant.
\newblock ``A method for determining the {{Zeeman}} splitting of a spin qubit
  via {{Rabi-driven}} tunneling''~(2025).
\newblock  \href{http://arxiv.org/abs/2503.17481}{arXiv:2503.17481}.

\bibitem{Elzerman2004}
J.~M. Elzerman, R.~Hanson, L.~H. {Willems van Beveren}, B.~Witkamp, L.~M.~K.
  Vandersypen, and L.~P. Kouwenhoven.
\newblock ``Single-shot read-out of an individual electron spin in a quantum
  dot''.
\newblock \href{https://dx.doi.org/10.1038/nature02693}{Nature {\bf 430},
  431--435}~(2004).

\bibitem{Koppens2006}
F.~H.~L. Koppens, C.~Buizert, K.~J. Tielrooij, I.~T. Vink, K.~C. Nowack,
  T.~Meunier, L.~P. Kouwenhoven, and L.~M.~K. Vandersypen.
\newblock ``Driven coherent oscillations of a single electron spin in a quantum
  dot''.
\newblock \href{https://dx.doi.org/10.1038/nature05065}{Nature {\bf 442},
  766--771}~(2006).

\bibitem{Lidar2020}
Daniel~A. Lidar.
\newblock ``Lecture {{Notes}} on the {{Theory}} of {{Open Quantum
  Systems}}''~(2020).
\newblock  \href{http://arxiv.org/abs/1902.00967}{arXiv:1902.00967}.

\bibitem{Rivas2012}
{\'A}ngel Rivas and Susana~F. Huelga.
\newblock ``Open quantum systems: An introduction''.
\newblock
  \href{https://dx.doi.org/10.1007/978-3-642-23354-8}{{{SpringerBriefs}} in
  {{Physics}}}. Springer-Verlag. Berlin Heidelberg~(2012).

\bibitem{Breuer2002}
Heinz-Peter Breuer and Francesco Petruccione.
\newblock ``The theory of open quantum systems''.
\newblock Oxford University Press. ~(2002).
\newblock  url:~\url{doi.org/10.1093/acprof:oso/9780199213900.001.0001}.

\bibitem{Lindblad1976}
G.~Lindblad.
\newblock ``On the generators of quantum dynamical semigroups''.
\newblock \href{https://dx.doi.org/10.1007/BF01608499}{Communications in
  Mathematical Physics {\bf 48}, 119--130}~(1976).

\bibitem{Gorini1976}
Vittorio Gorini, Andrzej Kossakowski, and E.~C.~G. Sudarshan.
\newblock ``Completely positive dynamical semigroups of {{N}}-level systems''.
\newblock \href{https://dx.doi.org/10.1063/1.522979}{Journal of Mathematical
  Physics {\bf 17}, 821--825}~(1976).

\bibitem{Martin2003}
I.~Martin, D.~Mozyrsky, and H.~W. Jiang.
\newblock ``A {{Scheme}} for {{Electrical Detection}} of {{Single-Electron Spin
  Resonance}}''.
\newblock \href{https://dx.doi.org/10.1103/PhysRevLett.90.018301}{Physical
  Review Letters {\bf 90}, 018301}~(2003).

\bibitem{Engel2001}
Hans-Andreas Engel and Daniel Loss.
\newblock ``Detection of {{Single Spin Decoherence}} in a {{Quantum Dot}} via
  {{Charge Currents}}''.
\newblock \href{https://dx.doi.org/10.1103/PhysRevLett.86.4648}{Physical Review
  Letters {\bf 86}, 4648--4651}~(2001).

\bibitem{Engel2002}
Hans-Andreas Engel and Daniel Loss.
\newblock ``Single-spin dynamics and decoherence in a quantum dot via charge
  transport''.
\newblock \href{https://dx.doi.org/10.1103/PhysRevB.65.195321}{Physical Review
  B {\bf 65}, 195321}~(2002).

\bibitem{Cohen-Tannoudji1977}
C.~{Cohen-Tannoudji} and S.~Reynaud.
\newblock ``Dressed-atom description of resonance fluorescence and absorption
  spectra of a multi-level atom in an intense laser beam''.
\newblock \href{https://dx.doi.org/10.1088/0022-3700/10/3/005}{Journal of
  Physics B: Atomic and Molecular Physics {\bf 10}, 345}~(1977).

\bibitem{Cohen-Tannoudji2008}
Claude {Cohen-Tannoudji}, Jacques {Dupont-Roc}, and Gilbert Grynberg, editors.
\newblock ``Atom-photon interactions: Basic processes and applications''.
\newblock \href{https://dx.doi.org/10.1002/9783527617197}{A
  {{Wiley-Interscience}} Publication}. Wiley. New York, NY~(2008).

\bibitem{Elouard2020}
Cyril Elouard, David {Herrera-Mart{\'i}}, Massimiliano Esposito, and Alexia
  Auff{\`e}ves.
\newblock ``Thermodynamics of optical {{Bloch}} equations''.
\newblock \href{https://dx.doi.org/10.1088/1367-2630/abbd6e}{New Journal of
  Physics {\bf 22}, 103039}~(2020).

\bibitem{Autler1955}
S.~H. Autler and C.~H. Townes.
\newblock ``Stark {{Effect}} in {{Rapidly Varying Fields}}''.
\newblock \href{https://dx.doi.org/10.1103/PhysRev.100.703}{Physical Review
  {\bf 100}, 703--722}~(1955).

\bibitem{Mozgunov2020}
Evgeny Mozgunov and Daniel Lidar.
\newblock ``Completely positive master equation for arbitrary driving and small
  level spacing''.
\newblock \href{https://dx.doi.org/10.22331/q-2020-02-06-227}{Quantum {\bf 4},
  227}~(2020).

\bibitem{Beenakker1991}
C.~W.~J. Beenakker.
\newblock ``Theory of {{Coulomb-blockade}} oscillations in the conductance of a
  quantum dot''.
\newblock \href{https://dx.doi.org/10.1103/PhysRevB.44.1646}{Physical Review B
  {\bf 44}, 1646--1656}~(1991).

\bibitem{Lidar2001}
Daniel~A. Lidar, Zsolt Bihary, and K.~Birgitta Whaley.
\newblock ``From completely positive maps to the quantum {{Markovian}}
  semigroup master equation''.
\newblock \href{https://dx.doi.org/10.1016/S0301-0104(01)00330-5}{Chemical
  Physics {\bf 268}, 35--53}~(2001).

\bibitem{supp}
``Supplemental {{Information}}''.

\bibitem{Dutra2004}
Sergio~M. Dutra.
\newblock ``Appendix {{F}}: {{The Good}}, the bad, and the {{Ugly}}:
  {{Principal Parts}}, {{Step}} and {{Delta Functions}}''.
\newblock In Cavity {{Quantum Electrodynamics}}.
\newblock \href{https://dx.doi.org/10.1002/0471713465.app6}{Pages 321--334}.
\newblock John Wiley \& Sons, Ltd~(2004).

\bibitem{Kasatkin2023}
Victor Kasatkin, Larry Gu, and Daniel~A. Lidar.
\newblock ``Which differential equations correspond to the {{Lindblad}}
  equation?''.
\newblock \href{https://dx.doi.org/10.1103/PhysRevResearch.5.043163}{Physical
  Review Research {\bf 5}, 043163}~(2023).

\bibitem{Hanson2007}
R.~Hanson, L.~P. Kouwenhoven, J.~R. Petta, S.~Tarucha, and L.~M.~K.
  Vandersypen.
\newblock ``Spins in few-electron quantum dots''.
\newblock \href{https://dx.doi.org/10.1103/RevModPhys.79.1217}{Reviews of
  Modern Physics {\bf 79}, 1217--1265}~(2007).

\bibitem{Baumann2015}
Susanne Baumann, William Paul, Taeyoung Choi, Christopher~P. Lutz, Arzhang
  Ardavan, and Andreas~J. Heinrich.
\newblock ``Electron paramagnetic resonance of individual atoms on a surface''.
\newblock \href{https://dx.doi.org/10.1126/science.aac8703}{Science {\bf 350},
  417--420}~(2015).

\bibitem{Ast2024}
Christian~R. Ast, Piotr Kot, Maneesha Ismail, Sebasti{\'a}n {de-la-Pe{\~n}a},
  Antonio~I. {Fern{\'a}ndez-Dom{\'i}nguez}, and Juan~Carlos Cuevas.
\newblock ``Theory of {{Electron Spin Resonance}} in {{Scanning Tunneling
  Microscopy}}''~(2024).
\newblock  \href{http://arxiv.org/abs/2403.20247}{arXiv:2403.20247}.

\bibitem{Mattuck1992}
Richard~D. Mattuck.
\newblock ``A {{Guide}} to {{Feynman Diagrams}} in the {{Many-Body Problem}}:
  {{Second Edition}}''.
\newblock Dover Publications. New York~(1992).
\newblock 2nd edition.
\newblock
  url:~\url{https://store.doverpublications.com/products/9780486670478}.

\end{thebibliography}

\end{document}